\documentclass[aps,prd,twocolumn,showpacs,superscriptaddress,groupedaddress]{revtex4-2}  
\usepackage{lineno} 
\usepackage{graphicx}  
\usepackage{dcolumn}   
\usepackage{bm}        
\usepackage{amssymb}   
\usepackage{xcolor}    
\usepackage{soul}
\usepackage{booktabs}  
\usepackage[shortlabels]{enumitem}

\begin{document}

\widetext


\title{Longitudinal double-spin asymmetry for inclusive jet and dijet production in polarized proton collisions at $\sqrt{s}=510$ GeV}
\author{
M.~S.~Abdallah$^{5}$,
B.~E.~Aboona$^{55}$,
J.~Adam$^{6}$,
L.~Adamczyk$^{2}$,
J.~R.~Adams$^{39}$,
J.~K.~Adkins$^{30}$,
G.~Agakishiev$^{28}$,
I.~Aggarwal$^{41}$,
M.~M.~Aggarwal$^{41}$,
Z.~Ahammed$^{60}$
% ...
}

\address{\rm{(STAR Collaboration)}}

% ...
\address{$^{2}$AGH University of Science and Technology, FPACS, Cracow 30-059, Poland}
% ...
\address{$^{5}$American University of Cairo, New Cairo 11835, New Cairo, Egypt}
\address{$^{6}$Brookhaven National Laboratory, Upton, New York 11973}
% ...
\address{$^{28}$Joint Institute for Nuclear Research, Dubna 141 980}
% ...
\address{$^{30}$University of Kentucky, Lexington, Kentucky 40506-0055}
% ...
\address{$^{39}$Ohio State University, Columbus, Ohio 43210}
% ...
\address{$^{41}$Panjab University, Chandigarh 160014, India}
% ...
\address{$^{55}$Texas A\&M University, College Station, Texas 77843}
% ...
\address{$^{60}$Variable Energy Cyclotron Centre, Kolkata 700064, India}
% ...
\collaboration{STAR Collaboration}
\date{\today}

\begin{abstract}
We report measurements of the longitudinal double-spin asymmetry, $A_{LL}$, for inclusive jet and dijet production in polarized proton-proton collisions at midrapidity and center-of-mass energy $\sqrt{s}$ = 510 GeV, using the high luminosity data sample collected by the STAR experiment in 2013. These measurements complement and improve the precision of previous STAR measurements at the same center-of-mass energy that probe the polarized gluon distribution function at partonic momentum fraction 0.015 $\lesssim x \lesssim$ 0.25. The dijet asymmetries are separated into four jet-pair topologies, which provide further constraints on the $x$ dependence of the polarized gluon distribution function. These measurements are in agreement with previous STAR measurements and with predictions from current next-to-leading-order global analyses. They provide more precise data at low dijet invariant mass that will better constrain the shape of the polarized gluon distribution function of the proton.
\end{abstract}

\maketitle

\section{\label{sec:Int}Introduction}

Over the last 20 years, the STAR experiment at the Relativistic Heavy Ion Collider (RHIC) has used high-energy polarized proton collisions with center-of-mass energies up to 510 GeV to gain deeper insight into the spin structure and dynamics of the proton. One of the major goals of the RHIC spin program is to perform high precision measurements of the polarized gluon distribution function of the proton, $\Delta g(x,Q^2)$, where $x$ is the partonic momentum fraction and $Q^2$ is the momentum transfer. These measurements are motivated by previous analyses from other experiments, starting from the results of polarized deep inelastic scattering experiments in the late 1980's, that showed the proton spin could not originate only from the quarks, thereby initiating experimental searches for the gluon contribution to the proton spin (see \cite{Aidala:2013} and references therein).

The kinematic coverage at STAR provides access to gluons through the quark-gluon and gluon-gluon scatterings which dominate particle production at low and medium values of transverse momentum at RHIC. Previous STAR longitudinal double-spin asymmetry ($A_{LL}$) measurements of inclusive jets with pseudorapidity $|\eta| <$ 1 \cite{JetsALL:2009} and dijets with $|\eta| <$ 0.8 \cite{DiJetsALL:2009}, from data collected during the year 2009 with center-of-mass energy of 200 GeV, strongly suggest a nonzero gluon polarization for $x > 0.05$. The latest global analysis fits, DSSV14 \cite{DSSV:2014} and NNPDFpol1.1 \cite{NNPDF:2014}, which include the 2009 STAR inclusive jet measurements \cite{JetsALL:2009}, extract a positive contribution to the proton spin coming from gluon spin; however, the uncertainty remains large for $x <$ 0.05. Previous STAR analyses of inclusive and dijet cross sections show good agreement with theoretical next-to-leading-order perturbative QCD calculations, motivating their use for $A_{LL}$ measurements \cite{JetsALL:2006,DiJetsALL:2009}. 

It has been suggested that dijet production should be an effective observable to extract the $x$ dependence of the gluon polarization, since dijets provide better constraints on the underlying kinematics, e.g., compared to inclusive observables \cite{Stratmann:2006}. At leading order, the dijet invariant mass is proportional to the square root of the product of the partonic momentum fractions, $M_{inv}=\sqrt{s x_1 x_2}$, and the pseudorapidity sum of the two jets is proportional to the logarithmic ratio of the $x$ values, $\eta_3$ + $\eta_4 \propto$ log($x_1/x_2$) \footnote{the kinematics of the initial partons and final jets are denoted by subscripts 1,2 and 3,4, respectively}. Measurements at both $\sqrt{s}$ = 200 GeV and 510 GeV provide broad kinematic coverage in $x$. The wide acceptance of the STAR detector permits reconstruction of dijet events with different topological configurations, i.e., different pseudorapidity combinations that probe symmetric ($x_1 = x_2$) and asymmetric ($x_1 < x_2$ or $x_1 > x_2$) partonic collisions.

STAR has also measured $A_{LL}$ for dijet production with one or both jets in $0.8 < \eta < 1.8$, using the data collected during 2009 at $\sqrt{s}=200$ GeV \cite{DiJetsEndCap:2009}. A reweighting study of the DSSV14 fit was performed using the 2009 STAR dijet data \cite{DiJetsALL:2009,DiJetsEndCap:2009}. The results of this reweighted fit had a clear impact on our understanding of the gluon polarization in the region of $x \gtrsim$ 0.2 \cite{DSSV:2019}. 

The first STAR inclusive jet and dijet $A_{LL}$ measurements in longitudinally polarized proton collisions at $\sqrt{s}$ = 510 GeV and midrapidity $|\eta| <$ 0.9 were performed using data recorded in 2012 \cite{JetsALL:2012}, presenting good agreement with previous results in the overlapping $x$ region. In 2015, STAR concluded the longitudinally polarized proton program with another $\sqrt{s}$ = 200 GeV dataset. The 2015 inclusive jet and dijet results at midrapidity \cite{JetsALL:2015} are consistent and have better precision than the previous measurements \cite{JetsALL:2009, DiJetsALL:2009}, providing further evidence of a positive gluon polarization for $x >$ 0.05. Both the 2012 and 2015 results will provide new constraints on the gluon polarization at 0.015 $\lesssim x \lesssim$ 0.25 and 0.05 $\lesssim x \lesssim$ 0.5, respectively, when they are included in future global analyses. Other measurements to constrain the gluon polarization include inclusive pion production by PHENIX at midrapidity \cite{PHENIX:2016,PHENIX:2020}, and by STAR at 2.65 $< \eta <$ 3.9 \cite{FMS:2018}, which provides sensitivity down to $x \sim$ 0.001.

In this paper, we report measurements of $A_{LL}$ for inclusive jet and dijet production at $\sqrt{s}$ = 510 GeV using the data recorded by STAR during 2013 in the region $|\eta| <$ 0.9. The luminosity was approximately 250 pb$^{-1}$, which is almost 3 times higher than the previous year. The longitudinal double-spin asymmetry $A_{LL}$ calculations follow the same procedure as \cite{JetsALL:2009, DiJetsALL:2009, JetsALL:2012, JetsALL:2015}:

\begin{equation}
A_{LL} = \frac{\Sigma_{runs} P_Y P_B (N^{++} - rN^{+-})}{\Sigma_{runs} P_Y^2 P_B^2 (N^{++} + rN^{+-})},
\end{equation}
where $P_{B}$ and $P_{Y}$ are the measured polarizations of the beams (denoted blue and yellow), $N^{++}$ and $N^{+-}$ are the jet or dijet yields for equal and opposite proton beam helicity configurations, and $r$ is the relative luminosity, which is the ratio of the luminosities for different helicity configurations of the colliding beams. The beam polarizations and the relative luminosities were reasonably constant during individual experimental runs, which were each about 30 min in length throughout a 7 to 8 h RHIC fill. The relative luminosity had a multimodal distribution that varied between 0.87 and 1.12 (average 1.002), depending on beam conditions e.g., polarization pattern and beam intensity. The polarizations of the beams were measured for each RHIC fill by a proton-carbon based Coulomb-nuclear interference polarimeter \cite{pC}, calibrated by using a polarized hydrogen gas-jet target \cite{Hjet}. The average polarizations were $P_B$ = 56\% and $P_Y$ = 54\%, with a 6.4\% relative uncertainty on the product of the beam polarizations \cite{polUnc}.

\section{\label{sec:Dec}Experiment and Jet reconstruction}

The main tracking device at STAR is a time projection chamber (TPC) in a 0.5 T solenoidal magnetic field. The TPC acceptance is $|\eta| \lesssim$ 1.3 and 2$\pi$ in the azimuthal angle ($\phi$) \cite{TPC}. The barrel electromagnetic calorimeter (BEMC) \cite{BEMC} and the endcap electromagnetic calorimeter (EEMC) \cite{EEMC} were used to trigger on jets and measure their electromagnetic constituents. The BEMC covers $|\eta| \le$ 1.0 and the EEMC 1.1 $< \eta <$ 2.0, both with full azimuthal coverage. The helicity-dependent relative luminosity was calculated using the vertex position detectors \cite{VPD} and zero degree calorimeters \cite{ZDC}, where the counts were corrected for accidental and multiple coincidences as in \cite{Coincidence:2000}.

Events were recorded if they satisfied a jet patch (JP) trigger condition \cite{JetsALL:2012,trigger}, which was defined by requiring that the BEMC or EEMC detected a transverse energy that exceeded one of the three thresholds equivalent to 6.8 GeV for JP0, 9.0 GeV for JP1 and 14.4 GeV for JP2, over an area of approximately $\Delta \eta \, \times \, \Delta \phi \, = 1 \times 1$. In addition to the JP triggers, two new triggers, ``JP0dijet" and ``JP1dijet", were introduced for this measurement. These new triggers required that one JP met the JP0 or JP1 energy threshold, and that a second JP met a threshold of 2.8 GeV, with the two JPs required to be nonadjacent in $\phi$. All JP2 events were collected while JP1dijet and JP0dijet were prescaled (one dijet per 3 and 12 triggered events, respectively). The JP1 and JP0 triggers were highly prescaled (around 1 in 40 and 200 triggered events, respectively) in order to reserve data acquisition bandwidth for the dijet triggers. 

The anti-$k_T$ algorithm \cite{antikt} and FastJet \cite{fastjet} package were used to reconstruct jets. The jet resolution parameter for this analysis was $R$ = 0.5, in contrast to the studies at $\sqrt{s}$ = 200 GeV that used $R$ = 0.6 \cite{JetsALL:2009, DiJetsALL:2009, JetsALL:2015}. This parameter was lowered for the 510 GeV measurements \cite{JetsALL:2012} to reduce sensitivity to underlying-event effects. The individual tracks and towers had to meet certain conditions, similar to the quality assurance requirements as in \cite{JetsALL:2012}. The tracks from the TPC that were used in the jet finding algorithm satisfied a transverse momentum $p_T>$ 0.2 GeV/$c$, had at least 12 hit points in the TPC with more than 51\% of the possible hits along the reconstructed track segment, were associated to a collision vertex located within $\pm$ 90 cm of the nominal interaction point, and followed a $p_T$-dependent distance of closest approach (DCA) to the vertex. The DCA requirements were: less than 2 cm for $p_T <$ 0.5 GeV/$c$, less than 1 cm for $p_T >$ 1.5 GeV/$c$, and linearly interpolated between these two points. The BEMC and EEMC towers were required to have a transverse energy of $E_T>$ 0.2 GeV. On average, charge hadrons deposit approximately 30\% of their energy in the calorimeters. The TPC reconstructs all charged particles so including the tower energy associated with charged particles would overestimate the jet momentum. If a track pointed to the tower, the track $p_T$ (multiplied by $c$) was subtracted from the tower $E_T$ to avoid double counting of particles which were fully reconstructed by both the TPC and calorimeters; the tower was not used in the jet reconstruction if the difference was less than zero. The fraction of jet energy detected in the calorimeters ($R_{EM}$) was required to be less than 0.95. There is a significant excess of 100\% neutral jet candidates. At low jet $p_T$ the excess arises from upstream beam-gas interactions, while those at higher $p_T$ mostly arise from cosmic ray showers.

For inclusive jets, only the JP0, JP1, and JP2 triggered events were considered. Software cuts in $p_T$ were applied above the trigger thresholds to JP0 = 7.0 GeV/$c$, JP1 = 9.6 GeV/$c$ and JP2 = 15.3 GeV/$c$, to reduce reconstruction bias near the hardware thresholds. The reconstructed jet axis was required to lie within the location of the JP that fired the trigger. Any jet containing a track with a reconstructed $p_T >$ 30 GeV/$c$ was rejected, since the TPC resolution degrades at these momenta. The summed $p_T$ of all reconstructed tracks in the jet was required to be larger than 0.5 GeV/$c$ to remove, for example, noncollision backgrounds. In cases where more than one jet in an event satisfied the selection criteria (approximately 5\% of jet events), only the two highest $p_T$ jets were taken. 

The dijet analysis only considered the two largest $p_T$ jets in an event. As in Ref. \cite{JetsALL:2012}, the dijet opening-angle and pseudorapidity cuts were $\Delta\phi >$ 120$^o$ and $|\Delta\eta| <$ 1.6, to remove jets arising from hard gluon emission and to avoid having both jets fall near the detector acceptance limits. An empirical $p_T$-matching condition required the ratio of the leading and away-side jet transverse momenta be $p_T^{leading}/p_T^{away}< 6 - (0.08 \times p_T^{max})$, where $p_T^{max}$ is the highest transverse momentum track in either jet, to remove fake jets \cite{JetsALL:2012}. An asymmetric $p_T$ cut was applied, requiring one jet to have $p_T >$ 7.0 GeV/$c$ while the other jet had $p_T >$ 5.0 GeV/$c$, to allow comparison with theoretical models \cite{DSSV:2014, NNPDF:2014, DSSV:2019}. The same software cuts in $p_T$ as for inclusive jets were applied. At least one of the jets needed to point to the location of the JP that fired the JP0, JP1 or JP2 trigger, whereas both jets needed to match the JP0dijet or JP1dijet trigger locations.

The individual jets in a dijet were separated into three pseudorapidity regions: forward $0.3< \eta <0.9$, central $-0.3< \eta <0.3$, and backward $-0.9< \eta <-0.3$. The $A_{LL}$ measurements for dijets are presented in four topology bins A-D (Table \ref{tab:topo}), as in \cite{JetsALL:2012}, which allows discrimination between symmetric and asymmetric collisions in terms of the partonic momentum fractions $x_1$ and $x_2$.

\begin{table}
    \centering
    \caption{The four dijet topology bins A-D.}
  	\begin{tabular}{c c c c c c c}
  	\hline
  	\hline
Bin & & & & $\eta_3$ and $\eta_4$ regions & & Physics description \\ \hline
A   & & & & 0.3 $< |\eta_{3,4}| <$ 0.9; $\eta_{3} \cdot \eta_{4} >$ 0 & & Forward-forward \\
B   & & & & $|\eta_{3,4}| <$ 0.3;  0.3 $< |\eta_{4,3}| <$ 0.9 & & Forward-central \\
C   & & & &  $|\eta_{3,4}| <$ 0.3 &   & Central-central     \\
D   & & & & 0.3 $< |\eta_{3,4}| <$ 0.9; $\eta_{3} \cdot \eta_{4} <$ 0 & & Forward-backward    \\
\hline
  \end{tabular}
    \label{tab:topo}
\end{table}

Inclusive jet and dijet observables were corrected for underlying-event (UE) contributions using the off-axis cone method as in \cite{JetsALL:2012, ALICE_UE:2014}. This correction also provides a statistical subtraction of the pileup. Inclusive jet or dijet events were rejected if the ratio of the underlying-event correction divided by the jet $p_T$ or dijet $M_{inv}$ was greater than 34\% and 36\%, respectively, as in \cite{JetsALL:2012,JetsALL:2015}, to ensure that the jet or dijet was not shifted by more than two bin intervals of $p_T$ or $M_{inv}$. 

\section{\label{sec:Sim}Embedded Simulation}

Simulation events were produced to quantify the detector response, connecting the jets at detector level to the initial partonic level. These simulated events were also used to estimate systematic uncertainties and apply a trigger bias correction. The simulations were produced using PYTHIA 6.4.28 \cite{pythia} with the Perugia 2012 tune 370 \cite{perugia}, reducing the PARP(90) parameter to 0.213 as in \cite{JetsALL:2012, JetsALL:2015}. This parameter controls the energy dependence of the low-$p_T$ cut for the underlying-event generation, thereby providing better agreement with STAR inclusive pion measurements \cite{pion:2006, pion:2012}. The full detector response was simulated with GEANT 3 \cite{geant}, with the STAR configuration in 2013. The simulated events were embedded into randomly selected bunch crossings from real data to mimic real beam background, pileup, and detector inefficiencies. No significant differences were seen when comparing jets in low and high luminosity runs from data and embedded simulation.

A trigger software simulator was used in the off-line processing to incorporate time-dependent pedestal variations and detector efficiencies. The trigger emulator classified simulation events using the same logic as the data triggering but without prescale factors, in order to match the data to the simulation. In the case that a jet satisfied all the conditions to be classified as JP1, this event could be recorded as a JP0 trigger in the data because of the prescale; however, it would be considered as a JP1 trigger in the analysis because the emulator promotes it. Similar considerations were made for the dijet triggers.

\begin{figure}
\includegraphics[width=\columnwidth]{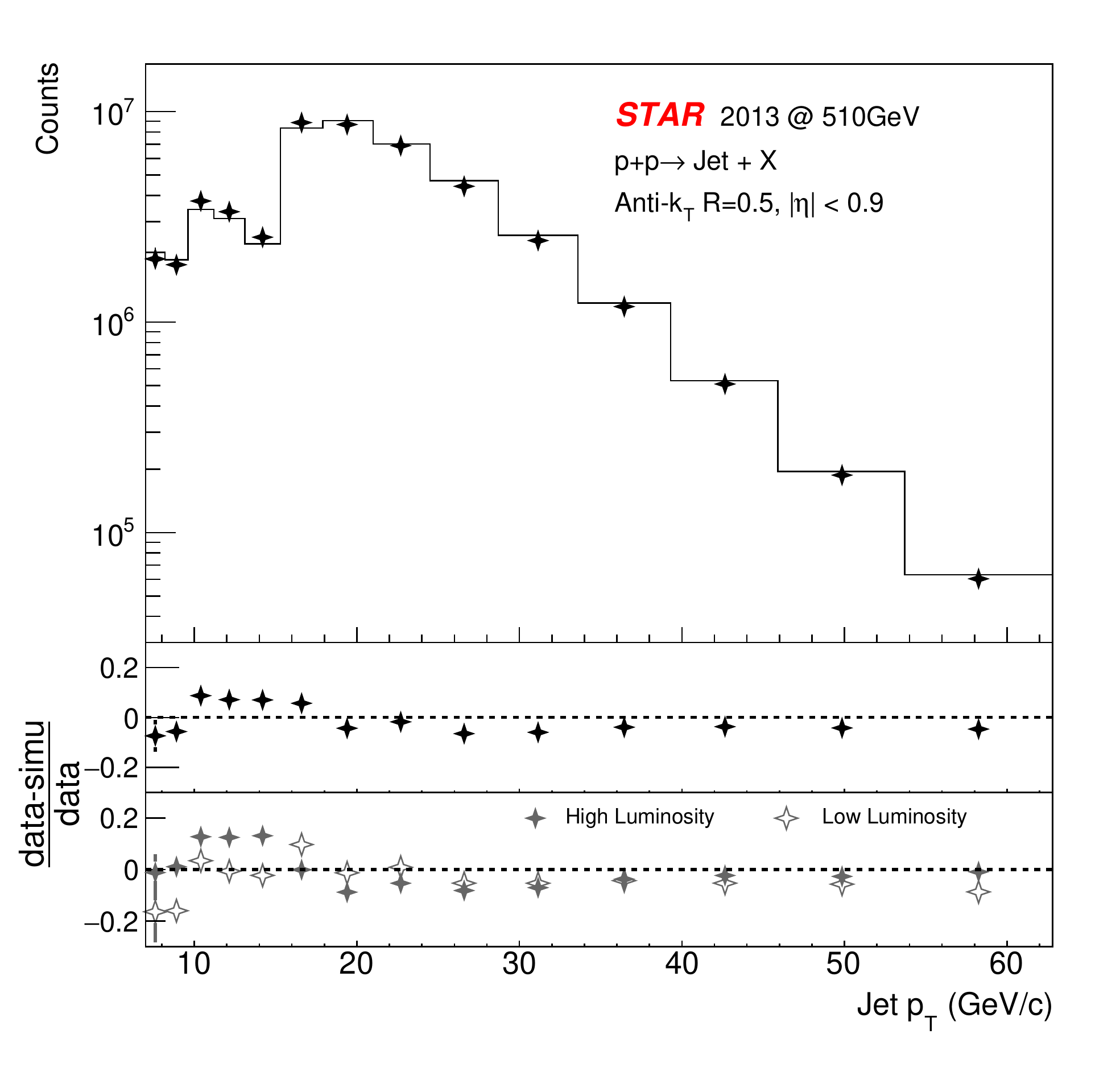}
\caption{\label{fig:zero} Comparison between data (points) and embedded simulation (histogram) of the inclusive jet yield versus $p_T$ at detector level. The central panel shows the ratio of the relative differences between all data runs used in the analysis and the simulation, and the lower panel shows the ratio for data separated into high and low luminosity runs. Statistical uncertainties are smaller than most of the points.}
\end{figure}

Figure \ref{fig:zero} shows the comparison between data and the embedded simulation of the inclusive jet counts versus $p_T$ at the detector level. The steps in the distribution correspond to trigger thresholds. A significant difference between 2012 and 2013 data is that much of the 2013 data were recorded under much higher instantaneous luminosity conditions. We verified that the embedded simulations provide comparable agreement with the data, independent of the instantaneous luminosity, as seen in the lower panel of Fig. \ref{fig:zero} which shows the ratio of data and simulation but only using high luminosity runs (approximately half of the full dataset) and low luminosity runs i.e. luminosity values comparable to 2012 data.  

\begin{figure}
\includegraphics[width=\columnwidth]{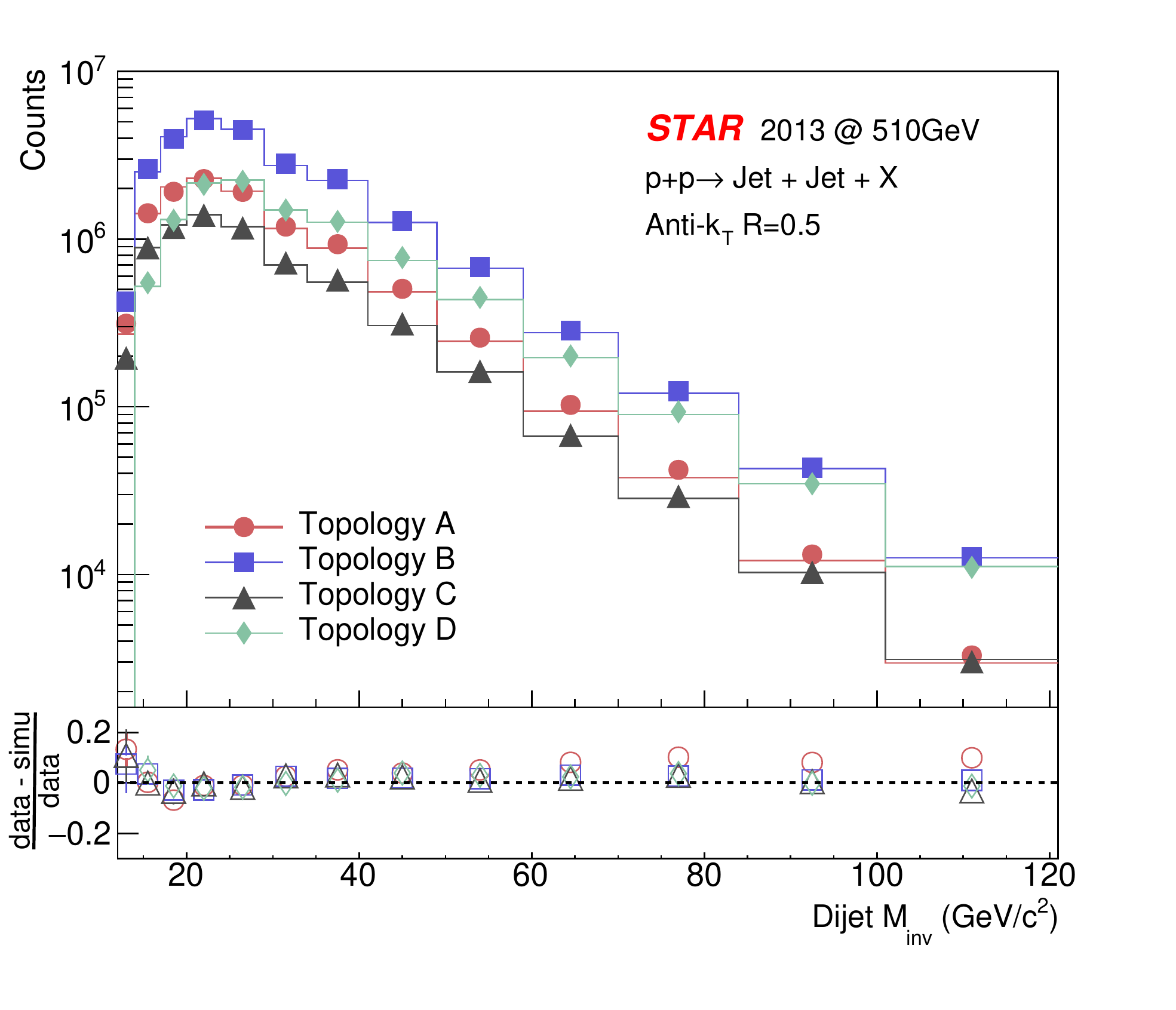}
\caption{\label{fig:one} Comparison between data (points) and embedded simulation (histograms) of the dijet yield versus the invariant mass, at detector level, for the different topology regions studied. The lower panel shows the ratio of the relative differences between data and the simulation. Statistical uncertainties are smaller than most of the points.}
\end{figure}

Figure \ref{fig:one} shows the comparison between data and embedded simulation of dijet counts versus invariant mass, at detector level, for the different topologies considered. The data and embedded simulation for both inclusive jet and dijet measurements agree to within 15\%; these differences are small enough to be covered by the systematic uncertainties. Data versus simulation comparisons were also examined for several other observables like: mean UE correction, $R_{EM}$, distributions of the charged hadrons within the jets as a function of the hadron longitudinal momentum fraction and as a function of the hadron momentum transverse to the thrust axis; and found to be comparable to the agreement of \cite{JetsALL:2012}.
 
The reconstructed jets were unfolded bin by bin to the parton level, in order to compare directly with the theoretical calculation, since $A_{LL}$ varies slowly and approximately linearly over the measured kinematic range. Higher-order distortions from resolution and efficiency, are accounted for as part of the trigger and reconstruction bias correction, which are found to be small. Detector jets were reconstructed in the simulation and matched with their partonic counterparts if the two jets were within $\sqrt{\Delta\eta^2+\Delta\phi^2} \le$ 0.5. The closest parton jet in $\eta - \phi$ space was chosen if more than one parton jet matched a given detector jet. The jet energy resolution was 100\%$/\sqrt{E}$ at low $p_{T}$, improving to 70\%$/\sqrt{E}$ at 20 GeV/$c$ $< p_{T} <$ 60 GeV/$c$.

\section{\label{sec:Sys}Systematic Uncertainties}

The sources of systematic uncertainties (syst. uncert.) considered for both the inclusive jet and dijet measurements are the same as in \cite{JetsALL:2012, JetsALL:2015}. The jet energy scale systematic uncertainties include the following:
\begin{enumerate}[(a)]
    \item The TPC tracking efficiency and resolution effects, or hadron response (Hadron resp.) were calculated by producing another simulation at the detector level and randomly rejecting 4\% of the reconstructed tracks to simulate a loss on track reconstruction efficiency as in \cite{JetsALL:2015}. The difference between this 4\% track loss sample and the nominal embedding sample is considered as a systematic uncertainty, with an additional 1\% component added in quadrature, associated with the efficiency of GEANT to simulate the EMCs response of charged hadrons.
    \item The electromagnetic response (EM resp.) was quantified by the quadrature sum of BEMC neutral energy uncertainty (5\%) plus the track efficiency for both TPC and BEMC (1\%). This was the dominant jet energy scale systematic.
    \item The difference between data and simulation for the underlying-event correction. 
    \item  The quadrature sum of the differences between other PYTHIA tunes and the nominal tune (370). At low $p_T$, the jet energy scale uncertainty is dominated not by PYTHIA uncertainties, but by the uncertainties in the calorimeter calibration and the ability of GEANT simulations to describe the interactions of hadrons. Differences between tunes are expected to be very small since some tunes vary the same set of parameters to control common activities and some tunes are related to the underlying event.
    \item The statistical uncertainties obtained after the bin by bin unfolding for $\delta p_T = \langle \delta p_{T,parton} - p_{T,detector} \rangle$ or $\delta M_{inv} = \langle \delta M_{inv,parton} - M_{inv,detector} \rangle$, due to the embedded simulation statistics.
\end{enumerate}

Table \ref{tab:ptSyst_incjet} summarizes the jet energy scale systematic uncertainties calculated for inclusive jets. Table \ref{tab:massSyst_dijet} summarizes the jet energy scale systematic uncertainties calculated for dijets in each topology. The last bins (101 - 121 GeV/$c^2$) for topologies A and C, and the first bin (12 - 14 GeV/$c^2$) of topology D, are not included due to poor statistics.

\begin{table*}
    \centering
    \caption{Jet energy scale correction and systematic uncertainties for inclusive jets. All values are given in GeV/$c$.}
    \begin{ruledtabular}
  	\begin{tabular}{c c c c c c c c c c}
    \toprule
    \multicolumn{1}{c}{ } & \multicolumn{3}{c}{Detector jet}  & \multicolumn{5}{c}{ } & Parton jet \\
    \cmidrule(lr){2-4} \cmidrule(lr){10-10}
    Bin label & $p_T$ range &	$R_{EM}$ & $\langle p_T \rangle$ &  $\delta p_T$ & Hadron resp. & EM resp. & UE syst. & Tune syst. & $p_T$ (final)\\
	\midrule
I1 & 7.0 – 8.2 &	0.67	 &	7.59	 &	0.20	 $\pm$ 0.21	 &	0.06	 &	0.27	 &	0.05	 & 0.78 &	7.79	 $\pm$	0.86	 \\
I2 & 8.2 – 9.6	 &	0.64	 &	8.81	 &	0.81	 $\pm$ 0.06	 &	0.07	 &	0.31	 &	0.01 &	0.49	 &	9.62	 $\pm$	0.59	 \\
I3 & 9.6 – 11.2 &	0.66	 &	10.35 &	1.32	 $\pm$ 0.05	 &	0.12	 &	0.37	 &	0.02 &	0.26	 &	11.67 $\pm$	0.47	 \\
I4 & 11.2 – 13.1 &	0.63	 &	12.04 &	1.55	 $\pm$ 0.05	 &	0.09	 &	0.42	 &	0.03 &	0.27	 &	13.59 $\pm$	0.51	 \\
I5 & 13.1 – 15.3	 &	0.58	 &	14.05 &	1.71	 $\pm$ 0.04	 &	0.18	 &	0.47	 &	0.04 &	0.19	 &	15.76 $\pm$	0.54	 \\
I6 & 15.3 – 17.9 &	0.72	 &	16.58 &	3.31	 $\pm$ 0.05	 &	0.19	 &	0.63	 &	0.10 &	0.28	 &	19.89 $\pm$	0.73	 \\
I7 & 17.9 – 20.9 &	0.69	 &	19.27 &	3.41	 $\pm$ 0.04	 &	0.13	 &	0.71	 &	0.09 &	0.35	 &	22.68 $\pm$	0.81	 \\
I8 & 20.9 – 24.5 &	0.64	 &	22.49 &	3.45	 $\pm$ 0.04	 &	0.20	 &	0.79	 &	0.09 &	0.35	 &	25.94 $\pm$	0.89	 \\
I9 & 24.5 – 28.7 &	0.59	 &	26.30 &	3.45	 $\pm$ 0.04	 &	0.19	 &	0.88	 &	0.09 &	0.43	 &	29.75 $\pm$	1.00	 \\
I10 & 28.7 – 33.6 &	0.55	 &	30.75 &	3.54	 $\pm$ 0.04	 &	0.26	 &	1.00	 &	0.11 &	0.63	 &	34.29 $\pm$	1.21	 \\
I11 & 33.6 – 39.3 &	0.52	 &	35.94 &	3.65	 $\pm$ 0.05	 &	0.30	 &	1.14	 &	0.11 &	0.74	 &	39.59 $\pm$	1.40	 \\
I12 & 39.3 – 46.0 &	0.51	 &	41.99 &	3.77	 $\pm$ 0.06	 &	0.26	 &	1.32	 &	0.11 &	0.70	 &	45.76 $\pm$	1.52	 \\
I13 & 46.0 – 53.8 &	0.50	 &	49.04 &	4.13	 $\pm$ 0.08	 &	0.34	 &	1.53	 &	0.11 &	0.71	 &	53.17 $\pm$	1.73	 \\
I14 & 53.8 – 62.8 &	0.51	 &	57.21 &	4.16	 $\pm$ 0.12	 &	0.27	 &	1.80	 &	0.10 &	0.68	 &	61.37 $\pm$	1.95	 \\
   \bottomrule
  \end{tabular}
  \end{ruledtabular}
    \label{tab:ptSyst_incjet}
\end{table*}

\begin{table*}
    \centering
    \caption{Jet energy scale correction and systematic uncertainties for the dijet topologies. All values are given in GeV/$c^2$.}
    \begin{ruledtabular}
  	\begin{tabular}{c c c c c c c c c c}
    \toprule
    \multicolumn{1}{c}{ } & \multicolumn{3}{c}{Detector jet}  & \multicolumn{5}{c}{ } & Parton jet \\
    \cmidrule(lr){2-4} \cmidrule(lr){10-10}
    Bin label & $M_{inv}$ range &	$R_{EM}$ & $\langle M_{inv} \rangle$ &  $\delta M_{inv}$ & Hadron resp. & EM resp. & UE syst. & Tune syst. & $M_{inv}$ (final) \\
	\midrule
	\multicolumn{10}{c}{Topology A: Forward-Forward Dijets}\\
A1 & 12 - 14	 &	0.58	 &	13.30	 &	2.44	 $\pm$	0.46	 &	0.61	 &	0.44	 &	0.07	 &	1.19	 &	15.74	 $\pm$	1.42	 \\
A2 & 14 - 17	 &	0.56	 &	15.63	 &	2.87	 $\pm$	0.20	 &	0.11	 &	0.51	 &	0.10	 &	0.57	 &	18.50	 $\pm$	1.14	 \\
A3 & 17 - 20	 &	0.55	 &	18.47	 &	3.49	 $\pm$	0.18	 &	0.40	 &	0.60	 &	0.08	 &	0.54	 &	21.96	 $\pm$	0.95	 \\
A4 & 20 - 24	 &	0.54	 &	21.84	 &	4.33	 $\pm$	0.13	 &	0.41	 &	0.70	 &	0.13	 &	0.44	 &	26.17	 $\pm$	0.94	 \\
A5 & 24 - 29	 &	0.52	 &	26.24	 &	5.39	 $\pm$	0.11	 &	0.47	 &	0.83	 &	0.13	 &	0.49	 &	31.63	 $\pm$	1.04	 \\
A6 & 29 - 34	 &	0.52	 &	31.24	 &	6.56	 $\pm$	0.12	 &	0.46	 &	0.99	 &	0.18	 &	0.83	 &	37.80	 $\pm$	1.36	 \\
A7 & 34 - 41	 &	0.51	 &	37.04	 &	7.71	 $\pm$	0.13	 &	0.54	 &	1.17	 &	0.17	 &	0.58	 &	44.75	 $\pm$	1.35	 \\
A8 & 41 - 49	 &	0.50	 &	44.40	 &	8.91	 $\pm$	0.14	 &	0.66	 &	1.39	 &	0.24	 &	0.42	 &	53.31	 $\pm$	1.50	 \\
A9 & 49 - 59	 &	0.49	 &	53.12	 &	10.52 $\pm$	0.17	 &	0.68	 &	1.65	 &  0.21	 &	0.35	 &	63.64	 $\pm$	1.73	 \\
A10 & 59 - 70	 &	0.48	 &	63.50	 &	11.82 $\pm$	0.24	 &	0.97	 &	1.96	 &	0.25	 &	0.53	 &	75.32	 $\pm$	2.06	 \\
A11 & 70 - 84	 &	0.47	 &	75.47	 &	14.19 $\pm$	0.28	 &	0.99	 &	2.32	 &	0.26	 &	0.89	 &	89.66	 $\pm$	2.51	 \\
A12 & 84 - 101	 &	0.46	 &	90.39	 &	15.24 $\pm$	0.40	 &	1.29	 &	2.76	 &	0.23	 &	0.79	 &	105.63 $\pm$ 2.90	 \\
A13 & 101 - 121 & - & - & - & - & - & - & - & - \\
    \multicolumn{10}{c}{Topology B: Forward-Central Dijets}\\
B1 & 12 - 14	 &	0.59	 &	13.38	 &	2.11	 $\pm$	0.70	 &	0.67	 &	0.45	 &	0.15	 &	1.00	 &	15.49	 $\pm$	1.47	 \\
B2 & 14 - 17	 &	0.58	 &	15.68	 &	2.80	 $\pm$	0.22	 &	0.22	 &	0.52	 &	0.06	 &	0.57	 &	18.48	 $\pm$	0.83	 \\
B3 & 17 - 20	 &	0.57	 &	18.49	 &	3.84	 $\pm$	0.13	 &	0.14	 &	0.61	 &	0.05	 &	0.39	 &	22.33	 $\pm$	0.75	 \\
B4 & 20 - 24	 &	0.55	 &	21.87	 &	4.80	 $\pm$	0.09	 &	0.09	 &	0.71	 &	0.12	 &	0.54	 &	26.67	 $\pm$	0.91	 \\
B5 & 24 - 29	 &	0.54	 &	26.25	 &	5.91	 $\pm$	0.08	 &	0.08	 &	0.84	 &	0.14	 &	0.48	 &	32.16	 $\pm$	0.99	 \\
B6 & 29 - 34	 &	0.53	 &	31.25	 &	7.34	 $\pm$	0.09	 &	0.09	 &	1.00	 &	0.20	 &	0.57	 &	38.59	 $\pm$	1.17	 \\
B7 & 34 - 41	 &	0.53	 &	37.06	 &	8.60	 $\pm$	0.08	 &	0.08	 &	1.18	 &	0.20	 &	0.73	 &	45.66	 $\pm$	1.41	 \\
B8 & 41 - 49	 &	0.52	 &	44.43	 &	10.06 $\pm$	0.10	 &	0.10	 &	1.41	 &	0.21	 &	0.49	 &	54.49	 $\pm$	1.51	 \\
B9 & 49 - 59	 &	0.50	 &	53.16	 &	11.86 $\pm$	0.12	 &	0.12	 &	1.66	 &	0.23	 &	0.45	 &	65.02	 $\pm$	1.75	 \\
B10 & 59 - 70	 &	0.50	 &	63.53	 &	13.63 $\pm$	0.16	 &	0.16	 &	1.99	 &	0.27	 &	0.46	 &	77.16	 $\pm$	2.07	 \\
B11 & 70 - 84	 &	0.49	 &	75.58	 &	15.75 $\pm$	0.19	 &	0.19	 &	2.35	 &	0.26	 &	0.97	 &	91.33	 $\pm$	2.57	 \\
B12 & 84 - 101	 &	0.49	 &	90.61	 &	18.59 $\pm$	0.25	 &	0.25	 &	2.82	 &	0.26	 &	1.09	 &	109.20$\pm$	3.05	 \\
B13 & 101 - 121 &	0.48	 &	108.55	 &	21.20 $\pm$	0.35	 &	0.35	 &	3.35	 &	0.23	 &	0.58	 &	129.75$\pm$	3.45	 \\
    \multicolumn{10}{c}{Topology C: Central-Central Dijets}\\
C1 & 12 - 14	 &	0.59	 &	13.30	 &	2.42	 $\pm$	0.72	 &	0.14	 &	0.45	 &	0.03	 &	1.28	 &	15.72	 $\pm$	1.54	 \\
C2 & 14 - 17	 &	0.58	 &	15.62	 &	3.28	 $\pm$	0.28	 &	0.10	 &	0.52	 &	0.07	 &	0.79	 &	18.90	 $\pm$	0.99	 \\
C3 & 17 - 20	 &	0.57	 &	18.47	 &	3.77	 $\pm$	0.38	 &	0.38	 &	0.61	 &	0.03	 &	0.61	 &	22.24	 $\pm$	1.02	 \\
C4 & 20 - 24	 &	0.56	 &	21.84	 &	5.30	 $\pm$	0.16	 &	0.33	 &	0.71	 &	0.17	 &	0.54	 &	27.14	 $\pm$	0.98	 \\
C5 & 24 - 29	 &	0.55	 &	26.24	 &	6.79	 $\pm$	0.13	 &	0.38	 &	0.85	 &	0.17	 &	0.52	 &	33.03	 $\pm$	1.09	 \\
C6 & 29 - 34	 &	0.54	 &	31.25	 &	8.10	 $\pm$	0.16	 &	0.46	 &	1.01	 &	0.20	 &	0.87	 &	39.35	 $\pm$	1.43	 \\
C7 & 34 - 41	 &	0.53	 &	37.04	 &	9.58	 $\pm$	0.16	 &	0.64	 &	1.18	 &	0.23	 &	0.70	 &	46.62	 $\pm$	1.54	 \\
C8 & 41 - 49	 &	0.52	 &	44.41	 &	11.33 $\pm$	0.18	 &	0.70	 &	1.41	 &	0.24	 &	0.60	 &	55.74	 $\pm$	1.71	 \\
C9 & 49 - 59	 &	0.51	 &	53.15	 &	13.24 $\pm$	0.22	 &	0.82	 &	1.67	 &	0.24	 &	0.40	 &	66.39	 $\pm$	1.93	 \\
C10 & 59 - 70	 &	0.50	 &	63.52	 &	15.45 $\pm$	0.29	 &	0.91	 &	1.99	 &	0.29	 &	0.40	 &	78.97	 $\pm$	2.26	 \\
C11 & 70 - 84	 &	0.50	 &	75.57	 &	18.65 $\pm$	0.35	 &	1.06	 &	2.36	 &	0.21	 &	1.10	 &	94.22	 $\pm$	2.85	 \\
C12 & 84 - 101	 &	0.49	 &	90.54	 &	20.94 $\pm$	0.48	 &	1.29	 &	2.81	 &	0.29	 &	0.70	 &	111.48 $\pm$	3.22	 \\
C13 & 101 - 121 & - & - & - & - & - & - & - & - \\
    \multicolumn{10}{c}{Topology D: Forward-Backward Dijets}\\
D1 & 12 - 14 & - & - & - & - & - & - & - & - \\
D2 & 14 - 17	 &	0.57	 &	15.87	 &	2.33	 $\pm$	0.37	 &	0.33	 &	0.52	 &	0.03	 &	0.90	 &	18.20	 $\pm$	1.15	 \\
D3 & 17 - 20	 &	0.56	 &	18.56	 &	3.11	 $\pm$	0.23	 &	0.36	 &	0.61	 &	0.05	 &	0.72	 &	21.67	 $\pm$	1.04	 \\
D4 & 20 - 24	 &	0.55	 &	21.96	 &	4.16	 $\pm$	0.16	 &	0.39	 &	0.71	 &	0.08	 &	0.62	 &	26.12	 $\pm$	1.04	 \\
D5 & 24 - 29	 &	0.53	 &	26.30	 &	5.12	 $\pm$	0.13	 &	0.41	 &	0.84	 &	0.07	 &	0.38	 &	31.42	 $\pm$	1.02	 \\
D6 & 29 - 34	 &	0.52	 &	31.27	 &	6.06	 $\pm$	0.14	 &	0.45	 &	0.99	 &	0.16	 &	0.57	 &	37.33	 $\pm$	1.25	 \\
D7 & 34 - 41	 &	0.51	 &	37.10	 &	7.50	 $\pm$	0.14	 &	0.51	 &	1.17	 &	0.16	 &	0.78	 &	44.60	 $\pm$	1.51	 \\
D8 & 41 - 49	 &	0.51	 &	44.48	 &	9.13	 $\pm$	0.15	 &	0.59	 &	1.40	 &	0.21	 &	0.69	 &	53.61	 $\pm$	1.69	 \\
D9 & 49 - 59	 &	0.50	 &	53.21	 &	10.17 $\pm$	0.18	 &	0.70	 &	1.66	 &	0.21	 &	0.62	 &	63.38	 $\pm$	1.93	 \\
D10 & 59 - 70	 &	0.49	 &	63.60	 &	11.92 $\pm$	0.22	 &	0.86	 &	1.98	 &	0.22	 &	0.66	 &	75.52	 $\pm$	2.27	 \\
D11 & 70 - 84	 &	0.48	 &	75.62	 &	13.98 $\pm$	0.28	 &	1.10	 &	2.34	 &	0.30	 &	0.65	 &	89.60	 $\pm$	2.70	 \\
D12 & 84 - 101	 &	0.47	 &	90.73	 &	15.90 $\pm$	0.37	 &	1.00	 &	2.79	 &	0.24	 &	1.13	 &	106.63 $\pm$	3.20	 \\
D13 & 101 - 121 &	0.47	 &	108.76	 &	18.41 $\pm$	0.48	 &	1.42	 &	3.34	 &	0.27	 &	1.07	 &	127.17 $\pm$	3.82	 \\
   \bottomrule
  \end{tabular}
  \end{ruledtabular}
    \label{tab:massSyst_dijet}
\end{table*}

Trigger and reconstruction bias effects were studied with the simulation to compensate for distortions due to detector finite resolution and efficiency. The efficiency of the STAR triggers varies for different partonic subprocesses (quark-quark, quark-gluon and gluon-gluon) \cite{JetsALL:2009, JetsALL:2012}. The trigger bias and the finite resolution of the detector affect the $A_{LL}$ measurements. Corrections were obtained by comparing the average differences between the asymmetry for reconstructed detector jets and parton jets, by using 100 equally probable replicas of the NNPDFpol1.1 \cite{NNPDF:2014} estimations. The root-mean-square of these differences (PDF uncert.), in addition to the finite statistics of the simulation (Stat. error), was considered as a systematic uncertainty.

The underlying-event correction modifies the value of the reconstructed jet energy, thus affecting the $A_{LL}$ measurement. Another systematic uncertainty was assigned to the $A_{LL}$ due to the underlying-event correction as in \cite{JetsALL:2012,JetsALL:2015}, by calculating the longitudinal double-spin asymmetry of the spin-dependent average underlying-event correction for inclusive jet, $A_{LL}^{dp_T}$, and dijet, $A_{LL}^{dM_{inv}}$. These underlying-event asymmetries were on average $A_{LL}^{dp_T}$ = 0.0006 $\pm$ 0.0009 for inclusive jet, $A_{LL}^{dM_{inv}}$ = -0.0006 $\pm$ 0.0010 for dijet topology A, -0.0001 $\pm$ 0.0007 for dijet topology B, -0.0015 $\pm$ 0.0013 for dijet topology C, and 0.0023 $\pm$ 0.0009 for dijet topology D.

The total $A_{LL}$ systematic uncertainties are the quadrature sum of the trigger and reconstruction bias, the underlying-event correction, plus the relative luminosity uncertainty that was estimated to be 4.7$\times 10^{-4}$. Tables \ref{tab:ALLSyst_incjet} and \ref{tab:ALLSyst_dijet} summarize the asymmetry corrections and systematic uncertainties calculated for inclusive jets and dijets in each topology. 

The parity-violating longitudinal single-spin asymmetries $A_L$ (for each of the two colliding beams) were consistent with zero within 2.5 standard deviations. The effect of a residual transverse beam polarization component was estimated and found to be negligible.

\begin{table*}
    \centering
    \caption{Asymmetry correction and systematic uncertainties for inclusive jets.}
    \begin{ruledtabular}
    \begin{tabular}{c c c c c c}
    \toprule  
           \hline
$p_T$ range (GeV/$c$) & Correction & PDF uncert. & Stat. error & UE syst. & Total $A_{LL}$ syst. \\
  	       \hline
7.0 – 8.2	 &	-0.00035	 &	0.00017	 &	0.00005	 &	0.00033	 &	0.00060	 \\
8.2 – 9.6	     &	-0.00034	 &	0.00013	 &	0.00006	 &	0.00027	 &	0.00056	 \\
9.6 – 11.2	 &	-0.00033	 &	0.00009	 &	0.00005	 &	0.00024	 &	0.00054	 \\
11.2 – 13.1	 &	-0.00038	 &	0.00010	 &	0.00007	 &	0.00020	 &	0.00053	 \\
13.1 – 15.3	 &	-0.00037	 &	0.00010	 &	0.00006	 &	0.00017	 &	0.00051	 \\
15.3 – 17.9	 &	-0.00013	 &	0.00007	 &	0.00006	 &	0.00016	 &	0.00050	 \\
17.9 – 20.9	 &	-0.00020	 &	0.00008	 &	0.00007	 &	0.00014	 &	0.00050	 \\
20.9 – 24.5	 &	-0.00016	 &	0.00014	 &	0.00009	 &	0.00012	 &	0.00051	 \\
24.5 – 28.7	 &	-0.00039	 &	0.00022	 &	0.00012	 &	0.00010	 &	0.00054	 \\
28.7 – 33.6	 &	-0.00020	 &	0.00031	 &	0.00017	 &	0.00009	 &	0.00059	 \\
33.6 – 39.3	 &	-0.00026	 &	0.00036	 &	0.00023	 &	0.00008	 &	0.00064	 \\
39.3 – 46.0	 &	-0.00025	 &	0.00034	 &	0.00034	 &	0.00007	 &	0.00068	 \\
46.0 – 53.8	 &	-0.00106	 &	0.00060	 &	0.00051	 &	0.00006	 &	0.00092	 \\
53.8 – 62.8	 &	-0.00018	 &	0.00190	 &	0.00081	 &	0.00006	 &	0.00212	 \\
           \hline         
    \end{tabular}
    \end{ruledtabular}
    \label{tab:ALLSyst_incjet}
\end{table*}

\begin{table*}
    \centering
    \caption{Asymmetry correction and systematic uncertainties for dijet topologies.}
    \begin{ruledtabular}
   \begin{tabular}{c c c c c c}
    \toprule
           \hline
$M_{inv}$ range (GeV/$c^2$) & Correction & PDF uncert. & Stat. error & UE syst. & Total $A_{LL}$ syst. \\
  	       \hline
  	       \multicolumn{6}{c}{Topology A: Forward-Forward Dijets}\\
12 - 14	 &	0.00008	 &	0.00013	 &	0.00010	 &	0.00023	 &	0.00055	 \\
14 - 17	 &	-0.00007	 &	0.00008	 &	0.00007	 &	0.00056	 &	0.00074	 \\
17 - 20	 &	-0.00018	 &	0.00011	 &	0.00010	 &	0.00006	 &	0.00050	 \\
20 - 24	 &	-0.00032	 &	0.00015	 &	0.00011	 &	0.00022	 &	0.00055	 \\
24 - 29	 &	-0.00036	 &	0.00021	 &	0.00013	 &	0.00019	 &	0.00056	 \\
29 - 34	 &	-0.00042	 &	0.00022	 &	0.00018	 &	0.00015	 &	0.00057	 \\
34 - 41	 &	-0.00046	 &	0.00025	 &	0.00023	 &	0.00013	 &	0.00059	 \\
41 - 49	 &	-0.00056	 &	0.00031	 &	0.00032	 &	0.00011	 &	0.00066	 \\
49 - 59	 &	0.00010	 &	0.00045	 &	0.00049	 &	0.00010	 &	0.00082	 \\
59 - 70	 &	-0.00104	 &	0.00065	 &	0.00080	 &	0.00008	 &	0.00114	 \\
70 - 84	 &	-0.00148	 &	0.00083	 &	0.00107	 &	0.00007	 &	0.00144	 \\
84 - 101	 &	-0.00115	 &	0.00084	 &	0.00173	 &	0.00007	 &	0.00198	 \\
101 - 121 & - & - & - & - & - \\
\multicolumn{6}{c}{Topology B: Forward-Central Dijets}\\
12 - 14	 &	-0.00005	 &	0.00008	 &	0.00007	 &	0.00004	 &	0.00048	 \\
14 - 17	 &	-0.00003	 &	0.00008	 &	0.00005	 &	0.00005	 &	0.00048	 \\
17 - 20	 &	-0.00007	 &	0.00009	 &	0.00007	 &	0.00001	 &	0.00048	 \\
20 - 24	 &	-0.00026	 &	0.00012	 &	0.00007	 &	0.00003	 &	0.00049	 \\
24 - 29	 &	-0.00041	 &	0.00017	 &	0.00008	 &	0.00003	 &	0.00051	 \\
29 - 34	 &	-0.00041	 &	0.00021	 &	0.00011	 &	0.00002	 &	0.00053	 \\
34 - 41	 &	-0.00047	 &	0.00026	 &	0.00014	 &	0.00002	 &	0.00056	 \\
41 - 49	 &	-0.00053	 &	0.00034	 &	0.00021	 &	0.00002	 &	0.00062	 \\
49 - 59	 &	-0.00109	 &	0.00048	 &	0.00030	 &	0.00001	 &	0.00074	 \\
59 - 70	 &	-0.00031	 &	0.00067	 &	0.00047	 &	0.00001	 &	0.00094	 \\
70 - 84	 &	-0.00082	 &	0.00073	 &	0.00065	 &	0.00001	 &	0.00108	 \\
84 - 101	 &	0.00063	 &	0.00060	 &	0.00094	 &	0.00001	 &	0.00121	 \\
101 - 121 &	-0.00162	 &	0.00232	 &	0.00149	 &	0.00001	 &	0.00280	 \\
\multicolumn{6}{c}{Topology C: Central-Central Dijets}\\
12 - 14	 &	0.00003	 &	0.00012	 &	0.00018	 &	0.00060	 &	0.00079	 \\
14 - 17	 &	0.00004	 &	0.00010	 &	0.00010	 &	0.00145	 &	0.00153	 \\
17 - 20	 &	-0.00011	 &	0.00012	 &	0.00011	 &	0.00015	 &	0.00052	 \\
20 - 24	 &	-0.00029	 &	0.00014	 &	0.00012	 &	0.00056	 &	0.00075	 \\
24 - 29	 &	-0.00021	 &	0.00018	 &	0.00016	 &	0.00047	 &	0.00071	 \\
29 - 34	 &	0.00023	 &	0.00025	 &	0.00023	 &	0.00040	 &	0.00070	 \\
34 - 41	 &	-0.00042	 &	0.00031	 &	0.00031	 &	0.00034	 &	0.00073	 \\
41 - 49	 &	-0.00044	 &	0.00042	 &	0.00044	 &	0.00029	 &	0.00082	 \\
49 - 59	 &	-0.00020	 &	0.00062	 &	0.00066	 &	0.00026	 &	0.00105	 \\
59 - 70	 &	0.00045	 &	0.00097	 &	0.00103	 &	0.00021	 &	0.00151	 \\
70 - 84	 &	0.00012	 &	0.00089	 &	0.00139	 &	0.00018	 &	0.00173	 \\
84 - 101	 &	-0.00297	 &	0.00105	 &	0.00205	 &	0.00015	 &	0.00236	 \\
101 - 121 & - & - & - & - & - \\
\multicolumn{6}{c}{Topology D: Forward-Backward Dijets}\\
12 - 14 & - & - & - & - & - \\
14 - 17	 &	-0.00014	 &	0.00012	 &	0.00008	 &	0.00110	 &	0.00120	 \\
17 - 20	 &	-0.00010	 &	0.00010	 &	0.00008	 &	0.00025	 &	0.00055	 \\
20 - 24	 &	-0.00037	 &	0.00013	 &	0.00009	 &	0.00102	 &	0.00113	 \\
24 - 29	 &	-0.00043	 &	0.00016	 &	0.00009	 &	0.00083	 &	0.00097	 \\
29 - 34	 &	-0.00055	 &	0.00020	 &	0.00013	 &	0.00068	 &	0.00086	 \\
34 - 41	 &	-0.00069	 &	0.00023	 &	0.00017	 &	0.00060	 &	0.00081	 \\
41 - 49	 &	-0.00068	 &	0.00027	 &	0.00025	 &	0.00054	 &	0.00080	 \\
49 - 59	 &	-0.00054	 &	0.00040	 &	0.00034	 &	0.00041	 &	0.00082	 \\
59 - 70	 &	-0.00119	 &	0.00053	 &	0.00051	 &	0.00036	 &	0.00094	 \\
70 - 84	 &	-0.00093	 &	0.00055	 &	0.00071	 &	0.00031	 &	0.00106	 \\
84 - 101	 &	-0.00063	 &	0.00042	 &	0.00102	 &	0.00028	 &	0.00123	 \\
101 - 121	 &	-0.00006	 &	0.00159	 &	0.00149	 &	0.00024	 &	0.00224	 \\
           \hline         
    \end{tabular}
    \end{ruledtabular}
    \label{tab:ALLSyst_dijet}
\end{table*}

\section{\label{sec:Res}Results}

\begin{figure}
\includegraphics[width=\columnwidth]{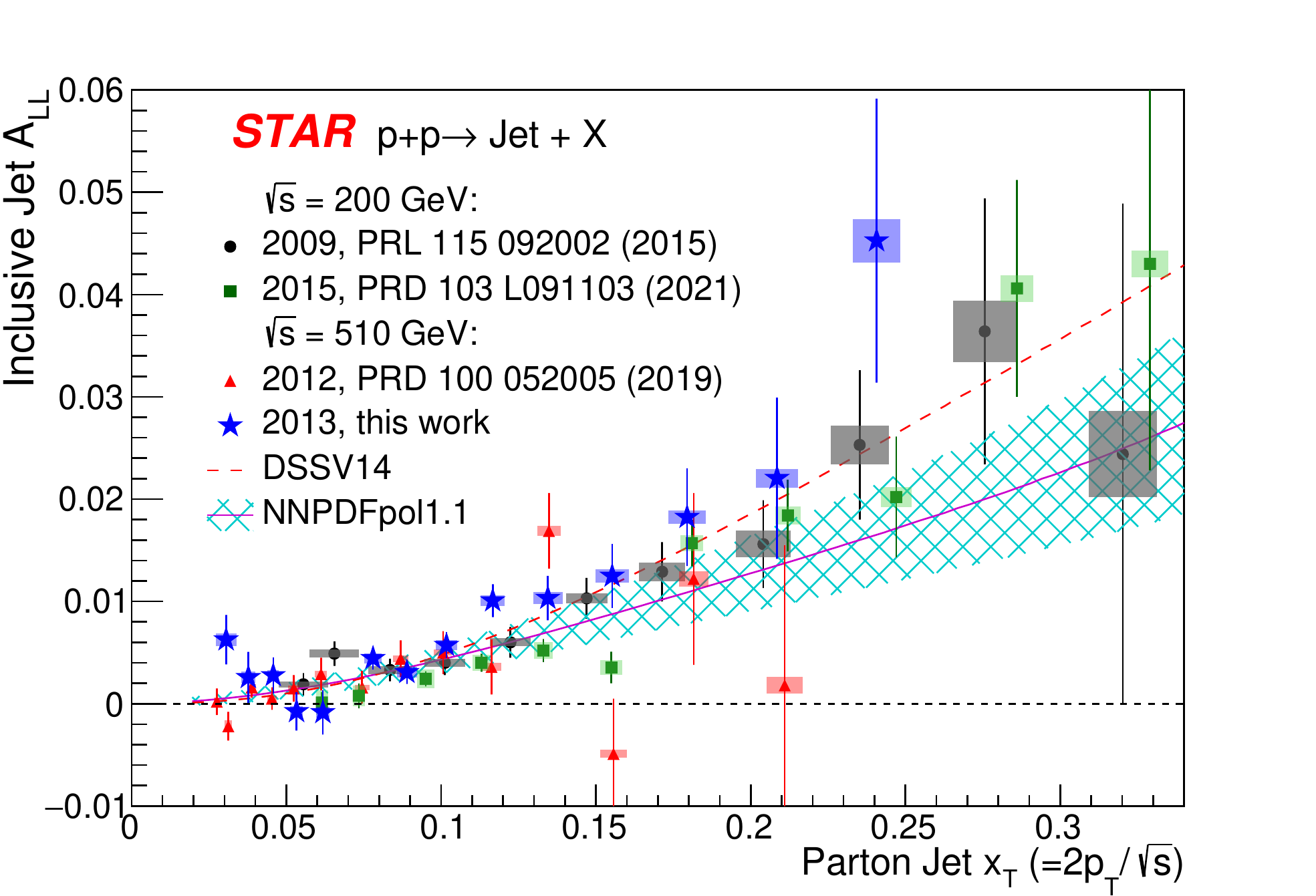}
\caption{\label{fig:three} Inclusive jet $A_{LL}$ versus $x_{T}$, compared to previous STAR results at $\sqrt{s}$ = 200 GeV \cite{JetsALL:2009,JetsALL:2015} and 510 GeV \cite{JetsALL:2012}, and evaluations from DSSV14 \cite{DSSV:2014} and NNPDFpol1.1 (with its uncertainty) \cite{NNPDF:2014} global analyses. The vertical lines are statistical uncertainties. The boxes show the size of the estimated systematic uncertainties. Scale uncertainties from polarization (not shown) are $\pm$6.5\%, $\pm$6.6\%, $\pm$6.4\% and $\pm$6.1\% from 2009 to 2015, respectively.}
\end{figure}

Figure \ref{fig:three} shows the 2013 inclusive jet $A_{LL}$ (blue) as a function of the parton jet transverse momentum scaled by 2/$\sqrt{s}$. The shaded blue boxes represent systematic uncertainty (width indicates the jet energy resolution).  The vertical lines correspond to statistical uncertainties, including consideration of the correlation between two jets when they are found in the same event. Table \ref{tab:final_incjet} presents the numerical results for the inclusive jet measurement. This result is compared with previous STAR results \cite{JetsALL:2009,JetsALL:2012,JetsALL:2015} with all their systematic uncertainties added in quadrature, and expectations from the latest global analyses available in \cite{DSSV:2014, NNPDF:2014}. There is good agreement among all measurements and with the global fits. 

\begin{table}
    \centering
    \caption{Inclusive jet $A_{LL}$ results.}
    \begin{ruledtabular}
   \begin{tabular}{c c c}
    \hline
    $p_T$ bin &	Jet $p_T$ & $A_{LL}$ $\pm$ stat. $\pm$ syst. \\
	\hline
7.0 – 8.2   &  7.79 $\pm$ 0.86 & 0.00626 $\pm$ 0.00241 $\pm$ 0.00060 \\
8.2 – 9.6    &  9.62 $\pm$ 0.59 & 0.00258 $\pm$ 0.00249 $\pm$ 0.00056 \\
9.6 – 11.2  & 11.67 $\pm$ 0.47 & 0.00277 $\pm$ 0.00176 $\pm$ 0.00054 \\
11.2 – 13.1 & 13.59 $\pm$ 0.51 &-0.00075 $\pm$ 0.00187 $\pm$ 0.00054 \\
13.1 – 15.3 & 15.76 $\pm$ 0.54 &-0.00085 $\pm$ 0.00216 $\pm$ 0.00051 \\
15.3 – 17.9 & 19.89 $\pm$ 0.73 & 0.00444 $\pm$ 0.00112 $\pm$ 0.00050 \\
17.9 – 20.9 & 22.68 $\pm$ 0.81 & 0.00308 $\pm$ 0.00114 $\pm$ 0.00050 \\
20.9 – 24.5 & 25.94 $\pm$ 0.89 & 0.00572 $\pm$ 0.00128 $\pm$ 0.00051 \\
24.5 – 28.7 & 29.75 $\pm$ 1.00 & 0.01008 $\pm$ 0.00161 $\pm$ 0.00054 \\
28.7 – 33.6 & 34.29 $\pm$ 1.21 & 0.01033 $\pm$ 0.00217 $\pm$ 0.00059 \\
33.6 – 39.3 & 39.59 $\pm$ 1.40 & 0.01249 $\pm$ 0.00312 $\pm$ 0.00064 \\
39.3 – 46.0 & 45.76 $\pm$ 1.52 & 0.01824 $\pm$ 0.00478 $\pm$ 0.00068 \\
46.0 – 53.8 & 53.17 $\pm$ 1.73 & 0.02205 $\pm$ 0.00788 $\pm$ 0.00092 \\
53.8 – 62.8 & 61.37 $\pm$ 1.95 & 0.04527 $\pm$ 0.01388 $\pm$ 0.00212 \\
   \hline
  \end{tabular}
  \end{ruledtabular}
    \label{tab:final_incjet}
\end{table}

\begin{figure}
\includegraphics[width=\columnwidth]{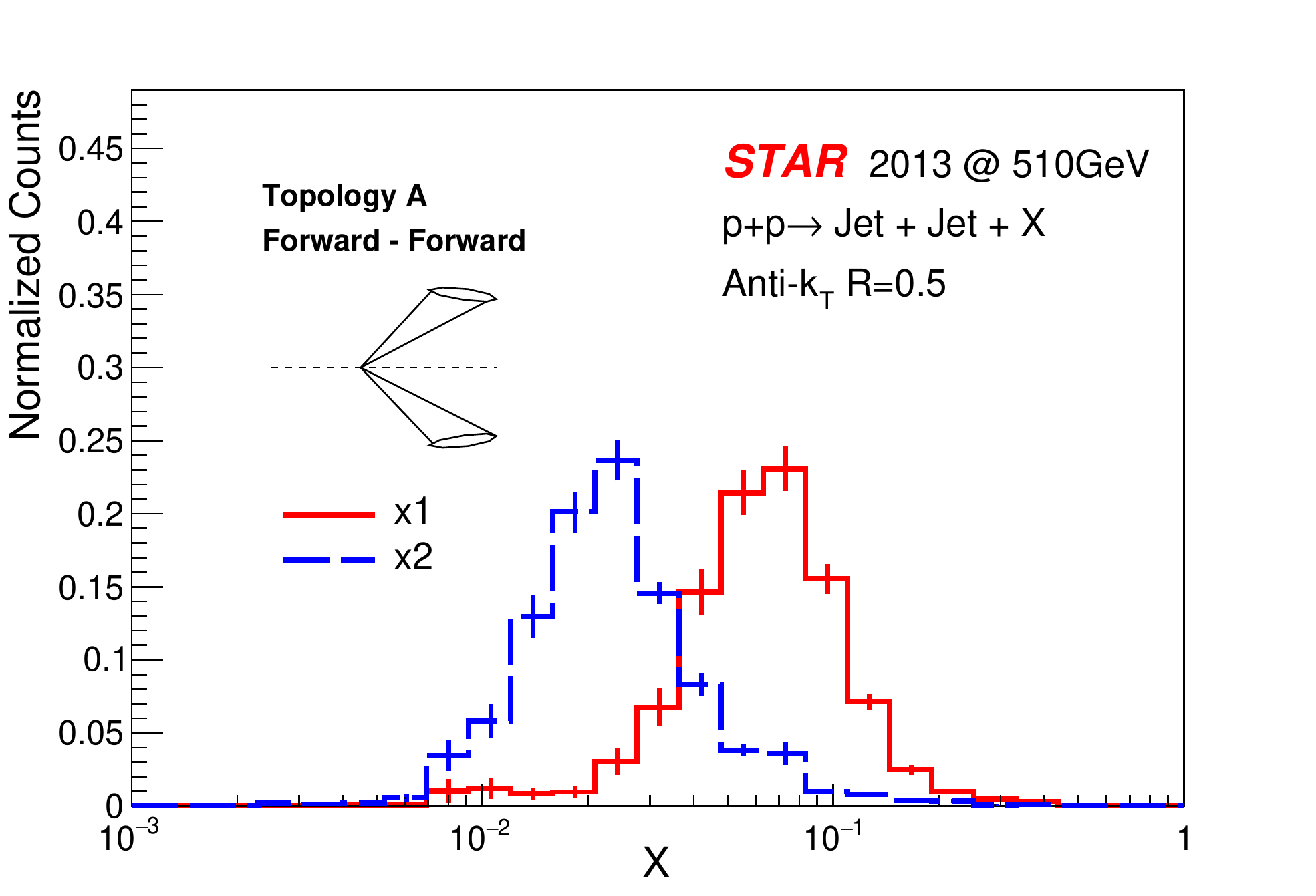}
\caption{\label{fig:two} Sampled $x_1$ (solid line) and $x_2$ (dashed line) gluon distributions weighted by the partonic asymmetry for dijet events with detector level $M_{inv}$ in the range 12 $< M_{inv} <$ 14 GeV/$c^{2}$, obtained using the embedded simulation for the topology A (the most asymmetric collisions). A representation of the topological configuration relative to the beam line is shown.}
\end{figure}

Figure \ref{fig:two} shows the $x_1$ and $x_2$ distributions using the reconstructed dijet events from the embedded simulation for the most asymmetric collisions (topology A) in the region 12 $< M_{inv} <$ 14 GeV/$c^{2}$. Figure \ref{fig:two} corresponds to the lowest momentum fraction values probed in these studies. The obtained values of $x_1$ and $x_2$ are weighted by the partonic asymmetry to indicate the region that is sensitive to the double-helicity measurement. The dijet triggers were introduced in this analysis specifically to enhance statistics at low $x$; sacrificing statistics at low $p_T$ for the inclusive jet measurement, as seen in Fig. \ref{fig:three}, while providing an order of magnitude greater statistics for the lower $M_{inv}$ bins for the dijet results. 

\begin{figure}[hb!]
\includegraphics[width=\columnwidth]{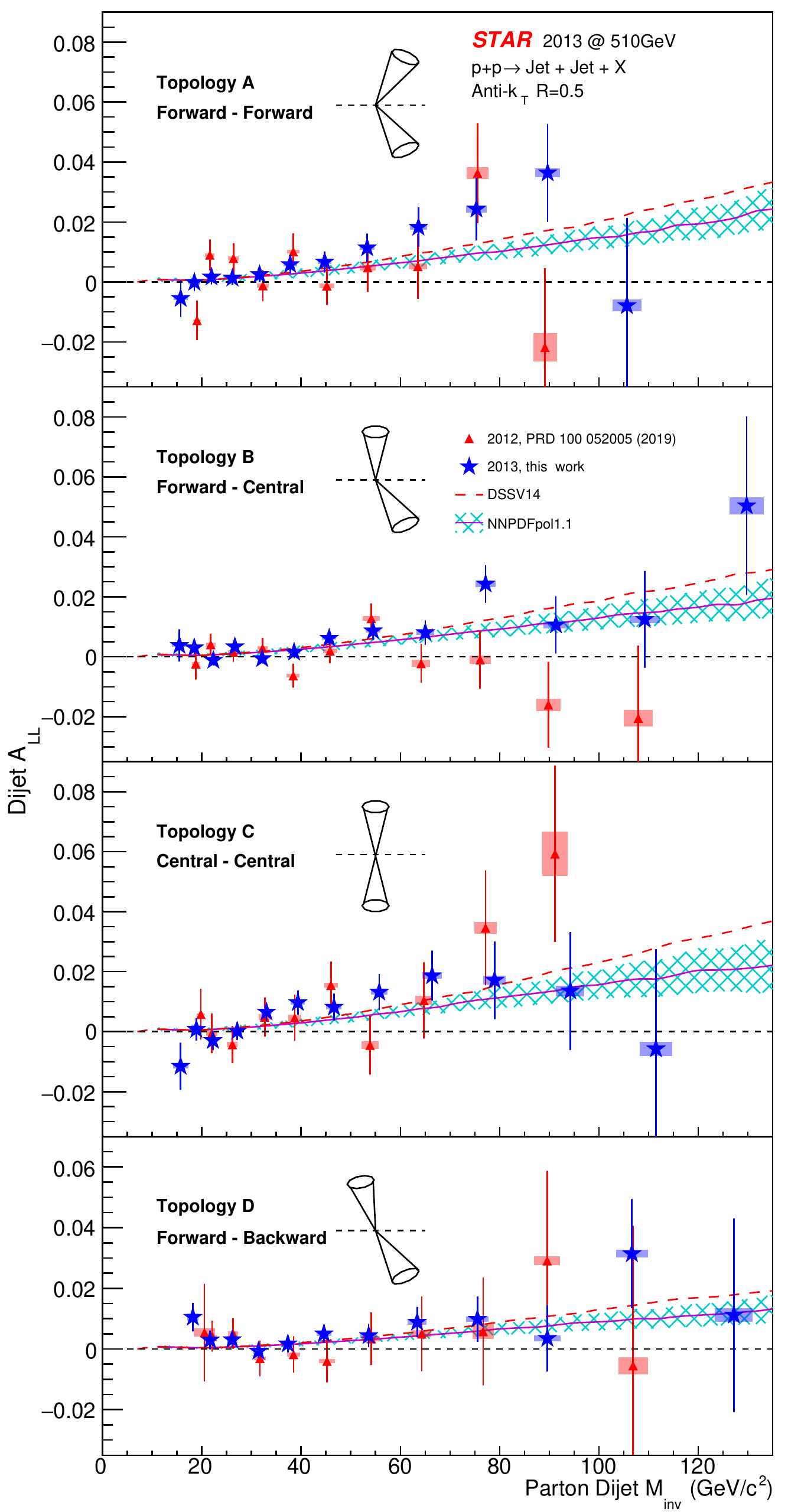}
\caption{\label{fig:four} Dijet $A_{LL}$ versus $M_{inv}$ for the A, B, C and D (top to bottom) topological configurations as explained in the text. They are compared to previous STAR results from 2012 data \cite{JetsALL:2012} and predictions from DSSV14 \cite{DSSV:2014} and NNPDFpol1.1 (with its uncertainty) \cite{NNPDF:2014} global analyses. The vertical lines are statistical uncertainties. The boxes show the size of the estimated systematic uncertainties. Topological configurations are shown for each jet orientation relative to the beam line. Scale uncertainties from polarization (not shown) are $\pm$6.6\% and $\pm$6.4\% for 2012 and 2013, respectively.}
\end{figure}

Figure \ref{fig:four} shows the dijet $A_{LL}$ as a function of the parton level invariant mass for the four topologies. Systematic uncertainties for dijet $A_{LL}$ were estimated following the same procedure as used for inclusive jet $A_{LL}$. The 2012 results \cite{JetsALL:2012} and the expectations from global analyses are also shown. Table \ref{tab:final_dijet} presents the numerical results of the dijet measurements in each topology. Similar to the inclusive jet results, there is good agreement between these and previous dijet results and with the global fits for all topologies.

\begin{table}
    \centering
    \caption{Dijet $A_{LL}$ results for each topology.}
    \begin{ruledtabular}
  	\begin{tabular}{c c c}
    \hline
    $M_{inv}$ bin &	Dijet $M_{inv}$ & $A_{LL}$ $\pm$ stat. $\pm$ syst. \\
	 \hline
\multicolumn{3}{c}{Topology A: Forward-Forward Dijets}\\
12 - 14 & 15.74 $\pm$ 1.42 &-0.00548 $\pm$ 0.00619 $\pm$ 0.00055 \\
14 - 17 & 18.50 $\pm$ 1.14 &-0.00011 $\pm$ 0.00289 $\pm$ 0.00074 \\
17 - 20 & 21.96 $\pm$ 0.95 & 0.00165 $\pm$ 0.00248 $\pm$ 0.00050 \\
20 - 24 & 26.17 $\pm$ 0.94 & 0.00129 $\pm$ 0.00226 $\pm$ 0.00055 \\
24 - 29 & 31.63 $\pm$ 1.04 & 0.00248 $\pm$ 0.00246 $\pm$ 0.00056 \\
29 - 34 & 37.80 $\pm$ 1.36 & 0.00581 $\pm$ 0.00311 $\pm$ 0.00057 \\
34 - 41 & 44.75 $\pm$ 1.35 & 0.00666 $\pm$ 0.00349 $\pm$ 0.00059 \\
41 - 49 & 53.31 $\pm$ 1.50 & 0.01140 $\pm$ 0.00472 $\pm$ 0.00066 \\
49 - 59 & 63.64 $\pm$ 1.73 & 0.01826 $\pm$ 0.00659 $\pm$ 0.00082 \\
59 - 70 & 75.32 $\pm$ 2.06 & 0.02431 $\pm$ 0.01045 $\pm$ 0.00114 \\
70 - 84 & 89.66 $\pm$ 2.51 & 0.03638 $\pm$ 0.01633 $\pm$ 0.00144 \\
84 - 101 & 105.63 $\pm$ 2.90 & -0.00789 $\pm$ 0.02919 $\pm$ 0.00198 \\
101 - 121 & - & - \\
\multicolumn{3}{c}{Topology B: Forward-Central Dijets}\\
12 - 14 & 15.49 $\pm$ 1.47 & 0.00381 $\pm$ 0.00533 $\pm$ 0.00048 \\
14 - 17 & 18.48 $\pm$ 0.83 & 0.00299 $\pm$ 0.00213 $\pm$ 0.00048 \\
17 - 20 & 22.33 $\pm$ 0.75 &-0.00116 $\pm$ 0.00173 $\pm$ 0.00048 \\
20 - 24 & 26.67 $\pm$ 0.91 & 0.00336 $\pm$ 0.00152 $\pm$ 0.00049 \\
24 - 29 & 32.16 $\pm$ 0.99 &-0.00060 $\pm$ 0.00161 $\pm$ 0.00051 \\
29 - 34 & 38.59 $\pm$ 1.17 & 0.00154 $\pm$ 0.00202 $\pm$ 0.00053 \\
34 - 41 & 45.66 $\pm$ 1.41 & 0.00620 $\pm$ 0.00224 $\pm$ 0.00056 \\
41 - 49 & 54.49 $\pm$ 1.51 & 0.00865 $\pm$ 0.00297 $\pm$ 0.00062 \\
49 - 59 & 65.02 $\pm$ 1.75 & 0.00806 $\pm$ 0.00406 $\pm$ 0.00074 \\
59 - 70 & 77.16 $\pm$ 2.07 & 0.02428 $\pm$ 0.00629 $\pm$ 0.00094 \\
70 - 84 & 91.33 $\pm$ 2.57 & 0.01063 $\pm$ 0.00953 $\pm$ 0.00108 \\
84 - 101 & 109.20 $\pm$ 3.05 & 0.01248 $\pm$ 0.01613 $\pm$ 0.00121 \\
101 - 121 & 129.75 $\pm$ 3.45 & 0.05037 $\pm$ 0.02978 $\pm$ 0.00280 \\
\multicolumn{3}{c}{Topology C: Central-Central Dijets}\\
12 - 14 & 15.72 $\pm$ 1.54 &-0.01155 $\pm$ 0.00785 $\pm$ 0.00079 \\
14 - 17 & 18.90 $\pm$ 0.99 & 0.00075 $\pm$ 0.00367 $\pm$ 0.00153 \\
17 - 20 & 22.24 $\pm$ 1.02 &-0.00293 $\pm$ 0.00317 $\pm$ 0.00052 \\
20 - 24 & 27.14 $\pm$ 0.98 & 0.00018 $\pm$ 0.00291 $\pm$ 0.00075 \\
24 - 29 & 33.03 $\pm$ 1.09 & 0.00655 $\pm$ 0.00317 $\pm$ 0.00071 \\
29 - 34 & 39.35 $\pm$ 1.43 & 0.00969 $\pm$ 0.00400 $\pm$ 0.00070 \\
34 - 41 & 46.62 $\pm$ 1.54 & 0.00817 $\pm$ 0.00448 $\pm$ 0.00073 \\
41 - 49 & 55.74 $\pm$ 1.71 & 0.01317 $\pm$ 0.00603 $\pm$ 0.00082 \\
49 - 59 & 66.39 $\pm$ 1.93 & 0.01866 $\pm$ 0.00833 $\pm$ 0.00105 \\
59 - 70 & 78.97 $\pm$ 2.26 & 0.01712 $\pm$ 0.01292 $\pm$ 0.00151 \\
70 - 84 & 94.22 $\pm$ 2.85 & 0.01357 $\pm$ 0.01964 $\pm$ 0.00173 \\
84 - 101 & 111.48 $\pm$ 3.22 & -0.00575 $\pm$ 0.03322 $\pm$ 0.00236 \\
101 - 121 & - & - \\
\multicolumn{3}{c}{Topology D: Forward-Backward Dijets}\\
12 - 14 & - & - \\
14 - 17 & 18.20 $\pm$ 1.15 & 0.01048 $\pm$ 0.00466 $\pm$ 0.00120 \\
17 - 20 & 21.67 $\pm$ 1.04 & 0.00296 $\pm$ 0.00302 $\pm$ 0.00055 \\
20 - 24 & 26.12 $\pm$ 1.04 & 0.00307 $\pm$ 0.00235 $\pm$ 0.00113 \\
24 - 29 & 31.42 $\pm$ 1.02 &-0.00072 $\pm$ 0.00229 $\pm$ 0.00097 \\
29 - 34 & 37.33 $\pm$ 1.25 & 0.00169 $\pm$ 0.00278 $\pm$ 0.00086 \\
34 - 41 & 44.60 $\pm$ 1.51 & 0.00501 $\pm$ 0.00300 $\pm$ 0.00081 \\
41 - 49 & 53.61 $\pm$ 1.69 & 0.00443 $\pm$ 0.00382 $\pm$ 0.00080 \\
49 - 59 & 63.38 $\pm$ 1.93 & 0.00887 $\pm$ 0.00500 $\pm$ 0.00082 \\
59 - 70 & 75.52 $\pm$ 2.27 & 0.00985 $\pm$ 0.00747 $\pm$ 0.00094 \\
70 - 84 & 89.60 $\pm$ 2.70 & 0.00351 $\pm$ 0.01095 $\pm$ 0.00106 \\
84 - 101 & 106.63 $\pm$ 3.20 & 0.03141 $\pm$ 0.01797 $\pm$ 0.00123 \\
101 - 121 & 127.17 $\pm$ 3.82 & 0.01114 $\pm$ 0.03187 $\pm$ 0.00224 \\
  \hline
  \end{tabular}
  \end{ruledtabular}
\label{tab:final_dijet}
\end{table}

There are point-to-point correlations between inclusive jet and dijet measurements from systematic effects, in addition to statistical correlations originating from the fact that $\sim$32\% of dijet events included at least one jet from the inclusive measurement. The underlying-event and trigger bias systematic uncertainties were treated as fully correlated in $A_{LL}$. Events with two reconstructed jets, both satisfying the inclusive jet conditions, will also introduce statistical correlations for the inclusive jet measurement. Total correlation matrices were calculated as in \cite{JetsALL:2012, JetsALL:2015} for inclusive-inclusive and inclusive-dijet events, and the systematic correlation matrices were calculated for dijet-dijet events (there is no statistical correlation for dijet $A_{LL}$). The relative luminosity uncertainty (4.7$\times 10^{-4}$) and the beam polarization uncertainty ($\pm$6.4\%), which are common to all the data points, were not included in the calculations. Correlation matrices are presented in the Appendix.  

\section{\label{sec:Con}Conclusions}

In summary, we report a high precision measurement of the inclusive jet and dijet longitudinal double-spin asymmetry $A_{LL}$ in polarized proton collisions at $\sqrt{s}$ = 510 GeV and $|\eta| <$ 0.9, using the large dataset collected by STAR in 2013. The results are consistent with previous STAR measurements and expectations from the latest global analyses, which included published RHIC data \cite{JetsALL:2009, DSSV:2014, NNPDF:2014}. The inclusive jet results will provide valuable new constraints on the magnitude of the gluon polarization, whereas the dijet results will have an impact on its functional form, in particular by using the topological configuration A that provides more precise data at low dijet invariant mass. These results provide sensitivity down to $x \sim$ 0.015, extending the kinematic coverage in future global analyses.

\begin{acknowledgments}

We thank the RHIC Operations Group and RCF at BNL, the NERSC Center at LBNL, and the Open Science Grid consortium for providing resources and support.  This work was supported in part by the Office of Nuclear Physics within the U.S. DOE Office of Science, the U.S. National Science Foundation, National Natural Science Foundation of China, Chinese Academy of Science, the Ministry of Science and Technology of China and the Chinese Ministry of Education, the Higher Education Sprout Project by Ministry of Education at NCKU, the National Research Foundation of Korea, Czech Science Foundation and Ministry of Education, Youth and Sports of the Czech Republic, Hungarian National Research, Development and Innovation Office, New National Excellency Programme of the Hungarian Ministry of Human Capacities, Department of Atomic Energy and Department of Science and Technology of the Government of India, the National Science Centre of Poland, the Ministry  of Science, Education and Sports of the Republic of Croatia, German Bundesministerium f\"ur Bildung, Wissenschaft, Forschung and Technologie (BMBF), Helmholtz Association, Ministry of Education, Culture, Sports, Science, and Technology (MEXT) and Japan Society for the Promotion of Science (JSPS).
\end{acknowledgments}

\appendix*
\section{\label{sec:App} CORRELATION MATRICES}
The inclusive jet and dijet results have two systematic uncertainties that are common to all the data points. The relative luminosity uncertainty represents a common offset of the $A_{LL} = 0$ axis by 4.7 x 10$^{-4}$. The product of the beam polarizations uncertainty is $\pm$ 6.4\%. In addition, there are point-to point statistical and systematic correlations previously discussed. The correlation matrix that quantifies these additional point-to-point effects is given in Tables \ref{tab:inc_inc}-\ref{tab:diD_diD}. The entries that are not shown can be obtained by transposition.

\begin{table*}
    \centering
     \caption{The correlation matrix for the point-to-point uncertainties (statistical and systematic) for the inclusive jet measurements. The relative luminosity and beam polarization uncertainties are not included because they are the same for all points.}
     \begin{ruledtabular}
  	\begin{tabular}{c | c c c c c c c c c c c c c c c c}
    \hline
    Labels & I1 & I2 & I3 & I4 & I5 & I6 & I7 & I8 & I9 & I10 & I11 & I12 & I13 & I14 \\
	\hline
  I1 &   1   &  0.007 &  0.004 &  0.004 &  0.003 &  0.001 &  0.001 &  0.000 &  0.000 &  0.000 &  0.000 &  0.000 &  0.000 &  0.000 \\
  I2 &       &    1   &  0.005 &  0.005 &  0.005 &  0.002 &  0.001 &  0.001 &  0.000 &  0.000 &  0.000 &  0.000 &  0.000 &  0.000 \\
  I3 &       &        &    1   &  0.008 &  0.009 &  0.004 &  0.002 &  0.002 &  0.001 &  0.001 &  0.000 &  0.000 &  0.000 &  0.000 \\
  I4 &       &        &        &    1   &  0.012 &  0.005 &  0.004 &  0.002 &  0.002 &  0.001 &  0.000 &  0.000 &  0.000 &  0.000 \\
  I5 &       &        &        &        &    1   &  0.007 &  0.005 &  0.004 &  0.003 &  0.002 &  0.001 &  0.000 &  0.000 &  0.000 \\
 I6 &       &        &        &        &        &    1   &  0.009 &  0.010 &  0.009 &  0.007 &  0.005 &  0.003 &  0.002 &  0.001 \\
 I7 &       &        &        &        &        &        &    1   &  0.013 &  0.013 &  0.011 &  0.009 &  0.005 &  0.003 &  0.001 \\
 I8 &       &        &        &        &        &        &        &    1   &  0.019 &  0.018 &  0.014 &  0.010 &  0.006 &  0.003 \\
 I9 &       &        &        &        &        &        &        &        &    1   &  0.024 &  0.022 &  0.016 &  0.010 &  0.005 \\
 I10 &       &        &        &        &        &        &        &        &        &    1   &  0.029 &  0.025 &  0.017 &  0.010 \\
 I11 &       &        &        &        &        &        &        &        &        &        &    1   &  0.033 &  0.026 &  0.017 \\
 I12 &       &        &        &        &        &        &        &        &        &        &        &    1   &  0.036 &  0.027 \\
 I13 &       &        &        &        &        &        &        &        &        &        &        &        &    1   &  0.037 \\
 I14 &       &        &        &        &        &        &        &        &        &        &        &        &        &    1   \\
   \hline
  \end{tabular}
  \end{ruledtabular}
    \label{tab:inc_inc}
\end{table*}


\begin{table*}
    \centering
    \caption{The correlation matrix for the point-to-point uncertainties (statistical and systematic) for the inclusive jet measurements coupling with the forward-forward dijet measurements (topology A). The relative luminosity and beam polarization uncertainties are not included because they are the same for all points.}
  	\begin{ruledtabular}
  	\begin{tabular}{c | c c c c c c c c c c c c c c c}
    \hline
    Label & A1 & A2 & A3 & A4 & A5 & A6 & A7 & A8 & A9 & A10 & A11 & A12 & A13 \\
	\hline
  I1 & 0.021  &  0.012  &  0.005  &  0.002  &  0.000  &  0.000  &  0.000  &  0.000  &  0.000  &  0.000  &  0.000  &  0.000  &  - \\
  I2 & 0.008  &  0.017  &  0.009  &  0.005  &  0.002  &  0.000  &  0.000  &  0.000  &  0.000  &  0.000  &  0.000  &  0.000  &  - \\
  I3 & 0.001  &  0.022  &  0.019  &  0.014  &  0.007  &  0.002  &  0.000  &  0.000  &  0.000  &  0.000  &  0.000  &  0.000  & - \\
  I4 & 0.000  &  0.010  &  0.021  &  0.020  &  0.015  &  0.006  &  0.002  &  0.000  &  0.000  &  0.000  &  0.000  &  0.000 & - \\
  I5 & 0.000  &  0.001  &  0.016  &  0.020  &  0.020  &  0.014  &  0.006  &  0.001  &  0.000  &  0.000  &  0.000  &  0.000  & - \\
 I6 & 0.000  &  0.000  &  0.013  &  0.033  &  0.047  &  0.049  &  0.041  &  0.017  &  0.004  &  0.000  &  0.000  &  0.000  &  - \\
 I7 & 0.000  &  0.000  &  0.002  &  0.023  &  0.041  &  0.055  &  0.066  &  0.045  &  0.015  &  0.002  &  0.000  &  0.000  &  - \\
 I8 & 0.000  &  0.000  &  0.000  &  0.010  &  0.027  &  0.043  &  0.071  &  0.077  &  0.047  &  0.012  &  0.001  &  0.000  &  - \\
 I9 & 0.000  &  0.000  &  0.000  &  0.002  &  0.014  &  0.025  &  0.051  &  0.081  &  0.088  &  0.043  &  0.009  &  0.001  &  - \\
 I10 & 0.000  &  0.000  &  0.000  &  0.000  &  0.004  &  0.011  &  0.026  &  0.055  &  0.094  &  0.090  &  0.039  &  0.006  &  - \\
 I11 & 0.000  &  0.000  &  0.000  &  0.000  &  0.000  &  0.003  &  0.011  &  0.027  &  0.062  &  0.101  &  0.094  &  0.030  &  - \\
 I12 & 0.000  &  0.000  &  0.000  &  0.000  &  0.000  &  0.000  &  0.003  &  0.010  &  0.029  &  0.066  &  0.114  &  0.091  &  - \\
 I13 & 0.000  &  0.000  &  0.000  &  0.000  &  0.000  &  0.000  &  0.000  &  0.002  &  0.010  &  0.028  &  0.071  &  0.116  &  - \\
 I14 & 0.000  &  0.000  &  0.000  &  0.000  &  0.000  &  0.000  &  0.000  &  0.000  &  0.002  &  0.009  &  0.028  &  0.076  &  - \\
 \hline	
	  \end{tabular}
	  \end{ruledtabular}
    \label{tab:inc_diA}
\end{table*}


\begin{table*}
    \centering
     \caption{The correlation matrix for the point-to-point uncertainties (statistical and systematic) for the inclusive jet measurements coupling with the forward-central dijet measurements (topology B). The relative luminosity and beam polarization uncertainties are not included because they are the same for all points.}
  	\begin{ruledtabular} 
  	\begin{tabular}{c | c c c c c c c c c c c c c c c}
    \hline
    Label & B1 & B2 & B3 & B4 & B5 & B6 & B7 & B8 & B9 & B10 & B11 & B12 & B13 \\
	\hline
  I1 & 0.026  &  0.022  &  0.009  &  0.004  &  0.001  &  0.000  &  0.000  &  0.000  &  0.000  &  0.000  &  0.000  &  0.000  &  0.000 \\
  I2 & 0.007  &  0.025  &  0.015  &  0.008  &  0.003  &  0.001  &  0.000  &  0.000  &  0.000  &  0.000  &  0.000  &  0.000  &  0.000 \\
  I3 & 0.001  &  0.025  &  0.029  &  0.022  &  0.013  &  0.005  &  0.001  &  0.000  &  0.000  &  0.000  &  0.000  &  0.000  &  0.000 \\
  I4 & 0.000  &  0.009  &  0.029  &  0.029  &  0.023  &  0.012  &  0.004  &  0.001  &  0.000  &  0.000  &  0.000  &  0.000  &  0.000 \\
  I5 & 0.000  &  0.001  &  0.017  &  0.027  &  0.029  &  0.022  &  0.012  &  0.003  &  0.000  &  0.000  &  0.000  &  0.000  &  0.000 \\
 I6 & 0.000  &  0.000  &  0.010  &  0.041  &  0.061  &  0.068  &  0.065  &  0.033  &  0.009  &  0.001  &  0.000  &  0.000  &  0.000 \\
 I7 & 0.000  &  0.000  &  0.001  &  0.025  &  0.052  &  0.073  &  0.095  &  0.075  &  0.033  &  0.007  &  0.001  &  0.000  &  0.000 \\
 I8 & 0.000  &  0.000  &  0.000  &  0.008  &  0.033  &  0.055  &  0.093  &  0.112  &  0.082  &  0.028  &  0.005  &  0.000  &  0.000 \\
 I9 & 0.000  &  0.000  &  0.000  &  0.001  &  0.015  &  0.031  &  0.063  &  0.105  &  0.129  &  0.080  &  0.025  &  0.003  &  0.000 \\
 I10 & 0.000  &  0.000  &  0.000  &  0.000  &  0.003  &  0.013  &  0.033  &  0.069  &  0.124  &  0.136  &  0.079  &  0.018  &  0.001 \\
 I11 & 0.000  &  0.000  &  0.000  &  0.000  &  0.000  &  0.003  &  0.013  &  0.033  &  0.079  &  0.136  &  0.149  &  0.069  &  0.011 \\
 I12 & 0.000  &  0.000  &  0.000  &  0.000  &  0.000  &  0.000  &  0.003  &  0.012  &  0.037  &  0.085  &  0.156  &  0.152  &  0.053 \\
 I13 & 0.000  &  0.000  &  0.000  &  0.000  &  0.000  &  0.000  &  0.000  &  0.002  &  0.012  &  0.038  &  0.096  &  0.173  &  0.144 \\
 I14 & 0.000  &  0.000  &  0.000  &  0.000  &  0.000  &  0.000  &  0.000  &  0.000  &  0.002  &  0.012  &  0.040  &  0.110  &  0.182 \\
 \hline
	  \end{tabular}\end{ruledtabular}
    \label{tab:inc_diB}
\end{table*}


\begin{table*}
    \centering
     \caption{The correlation matrix for the point-to-point uncertainties (statistical and systematic) for the inclusive jet measurements coupling with the central-central dijet measurements (topology C). The relative luminosity and beam polarization uncertainties are not included because they are the same for all points.}
  	\begin{ruledtabular} 
  	\begin{tabular}{c | c c c c c c c c c c c c c c c}
    \hline
    Label & C1 & C2 & C3 & C4 & C5 & C6 & C7 & C8 & C9 & C10 & C11 & C12 & C13 \\
	\hline
 I1 & 0.015  &  0.009  &  0.004  &  0.002  &  0.000  &  0.000  &  0.000  &  0.000  &  0.000  &  0.000  &  0.000  &  0.000  & - \\
  I2 & 0.006  &  0.012  &  0.007  &  0.004  &  0.001  &  0.000  &  0.000  &  0.000  &  0.000  &  0.000  &  0.000  &  0.000  & - \\
  I3 & 0.001  &  0.015  &  0.014  &  0.011  &  0.006  &  0.002  &  0.000  &  0.000  &  0.000  &  0.000  &  0.000  &  0.000  & - \\
  I4 & 0.000  &  0.007  &  0.016  &  0.015  &  0.012  &  0.005  &  0.002  &  0.000  &  0.000  &  0.000  &  0.000  &  0.000 & - \\
  I5 & 0.000  &  0.001  &  0.011  &  0.015  &  0.016  &  0.011  &  0.005  &  0.001  &  0.000  &  0.000  &  0.000  &  0.000  & - \\
 I6 & 0.000  &  0.000  &  0.008  &  0.023  &  0.033  &  0.035  &  0.030  &  0.013  &  0.003  &  0.000  &  0.000  &  0.000  & - \\
 I7 & 0.000  &  0.000  &  0.001  &  0.016  &  0.030  &  0.041  &  0.051  &  0.034  &  0.013  &  0.002  &  0.000  &  0.000  & - \\
 I8 & 0.000  &  0.000  &  0.000  &  0.007  &  0.020  &  0.033  &  0.054  &  0.059  &  0.036  &  0.011  &  0.002  &  0.000  & - \\
 I9 & 0.000  &  0.000  &  0.000  &  0.001  &  0.009  &  0.018  &  0.038  &  0.061  &  0.067  &  0.034  &  0.009  &  0.001  & - \\
 I10 & 0.000  &  0.000  &  0.000  &  0.000  &  0.003  &  0.008  &  0.020  &  0.042  &  0.072  &  0.070  &  0.033  &  0.006  & - \\
 I11 & 0.000  &  0.000  &  0.000  &  0.000  &  0.000  &  0.002  &  0.008  &  0.021  &  0.049  &  0.079  &  0.075  &  0.027  & - \\
 I12 & 0.000  &  0.000  &  0.000  &  0.000  &  0.000  &  0.000  &  0.002  &  0.008  &  0.024  &  0.054  &  0.090  &  0.074  & - \\
 I13 & 0.000  &  0.000  &  0.000  &  0.000  &  0.000  &  0.000  &  0.000  &  0.002  &  0.009  &  0.025  &  0.062  &  0.097  & - \\
 I14 & 0.000  &  0.000  &  0.000  &  0.000  &  0.000  &  0.000  &  0.000  &  0.000  &  0.002  &  0.008  &  0.028  &  0.073  & - \\
 \hline
	  \end{tabular}
	  \end{ruledtabular}
    \label{tab:inc_diC}
\end{table*}


\begin{table*}
    \centering
    \caption{The correlation matrix for the point-to-point uncertainties (statistical and systematic) for the inclusive jet measurements coupling with the forward-backward dijet measurements (topology D). The relative luminosity and beam polarization uncertainties are not included because they are the same for all points.}
    \begin{ruledtabular} 
    \begin{tabular}{c | c c c c c c c c c c c c c c c}
    \hline
    Label & D1 & D2 & D3 & D4 & D5 & D6 & D7 & D8 & D9 & D10 & D11 & D12 & D13 \\
	\hline
 I1 & -  &  0.019  &  0.010  &  0.004  &  0.002  &  0.000  &  0.000  &  0.000  &  0.000  &  0.000  &  0.000  &  0.000  &  0.000 \\
 I2 & -  &  0.013  &  0.014  &  0.008  &  0.004  &  0.001  &  0.000  &  0.000  &  0.000  &  0.000  &  0.000  &  0.000  &  0.000 \\
 I3 & -  &  0.006  &  0.021  &  0.017  &  0.012  &  0.006  &  0.002  &  0.000  &  0.000  &  0.000  &  0.000  &  0.000  &  0.000 \\
 I4 & -  &  0.001  &  0.013  &  0.019  &  0.017  &  0.012  &  0.006  &  0.001  &  0.000  &  0.000  &  0.000  &  0.000  &  0.000 \\
 I5 & -  &  0.000  &  0.004  &  0.015  &  0.019  &  0.017  &  0.013  &  0.005  &  0.001  &  0.000  &  0.000  &  0.000  &  0.000 \\
 I6 & -  &  0.000  &  0.001  &  0.016  &  0.033  &  0.042  &  0.050  &  0.036  &  0.015  &  0.003  &  0.000  &  0.000  &  0.000 \\
 I7 & -  &  0.000  &  0.000  &  0.006  &  0.025  &  0.038  &  0.058  &  0.060  &  0.039  &  0.013  &  0.002  &  0.000  &  0.000 \\
 I8 & -  &  0.000  &  0.000  &  0.001  &  0.013  &  0.026  &  0.048  &  0.069  &  0.071  &  0.038  &  0.011  &  0.001  &  0.000 \\
 I9 & -  &  0.000  &  0.000  &  0.000  &  0.004  &  0.014  &  0.029  &  0.053  &  0.081  &  0.076  &  0.038  &  0.008  &  0.000 \\
 I10 & -  &  0.000  &  0.000  &  0.000  &  0.000  &  0.005  &  0.014  &  0.030  &  0.061  &  0.089  &  0.082  &  0.031  &  0.005 \\
 I11 & -  &  0.000  &  0.000  &  0.000  &  0.000  &  0.000  &  0.004  &  0.013  &  0.033  &  0.066  &  0.103  &  0.080  &  0.025 \\
 I12 & -  &  0.000  &  0.000  &  0.000  &  0.000  &  0.000  &  0.001  &  0.004  &  0.013  &  0.035  &  0.076  &  0.111  &  0.073 \\
 I13 & -  &  0.000  &  0.000  &  0.000  &  0.000  &  0.000  &  0.000  &  0.000  &  0.004  &  0.013  &  0.038  &  0.088  &  0.114 \\
 I14 & -  &  0.000  &  0.000  &  0.000  &  0.000  &  0.000  &  0.000  &  0.000  &  0.000  &  0.003  &  0.013  &  0.040  &  0.096 \\
 \hline
	  \end{tabular}
	  \end{ruledtabular}
    \label{tab:inc_diD}
\end{table*}


\begin{table*}
    \centering
    \caption{The correlation matrix for the point-to-point uncertainties (systematic only) for forward-forward dijet measurements (topology A). The relative luminosity and beam polarization uncertainties are not included because they are the same for all points.}
  	\begin{ruledtabular} 
  	\begin{tabular}{c | c c c c c c c c c c c c c c c}
    \hline
    Label & A1 & A2 & A3 & A4 & A5 & A6 & A7 & A8 & A9 & A10 & A11 & A12 & A13 \\
	\hline
A1 &   1   &  0.016 &  0.019 &  0.021 &  0.019 &  0.015 &  0.014 &  0.010 &  0.007 &  0.005 &  0.003 &  0.002 &  - \\
A2 &       &    1   &  0.073 &  0.079 &  0.073 &  0.058 &  0.052 &  0.039 &  0.028 &  0.017 &  0.011 &  0.006 &  - \\
A3 &       &        &    1   &  0.042 &  0.039 &  0.031 &  0.028 &  0.021 &  0.015 &  0.009 &  0.006 &  0.003 &  - \\
A4 &       &        &        &    1   &  0.052 &  0.041 &  0.037 &  0.027 &  0.020 &  0.012 &  0.008 &  0.004 &  - \\
A5 &       &        &        &        &    1   &  0.039 &  0.035 &  0.026 &  0.019 &  0.012 &  0.008 &  0.004 &  - \\
A6 &       &        &        &        &        &    1   &  0.029 &  0.022 &  0.015 &  0.010 &  0.006 &  0.004 &  - \\
A7 &       &        &        &        &        &        &    1   &  0.021 &  0.015 &  0.009 &  0.006 &  0.003 &  - \\
A8 &       &        &        &        &        &        &        &    1   &  0.013 &  0.008 &  0.005 &  0.003 &  - \\
A9 &       &        &        &        &        &        &        &        &    1   &  0.010 &  0.006 &  0.003 &  - \\
A10 &       &        &        &        &        &        &        &        &        &    1   &  0.007 &  0.004 & - \\
A11 &       &        &        &        &        &        &        &        &        &        &    1   &  0.004 & - \\
A12 &       &        &        &        &        &        &        &        &        &        &        &    1   & - \\
A13 &       &        &        &        &        &        &        &        &        &        &        &        & -  \\
 \hline	
	  \end{tabular}
	  \end{ruledtabular}
    \label{tab:diA_diA}
\end{table*}


\begin{table*}
    \centering
    \caption{The correlation matrix for the point-to-point uncertainties (systematic only) coupling forward-forward dijet measurements (topology A) with forward-central dijet measurements (topology B). The relative luminosity and beam polarization uncertainties are not included because they are the same for all points.}
     \begin{ruledtabular} 
     \begin{tabular}{c | c c c c c c c c c c c c c c c}
    \hline
    Label & B1 & B2 & B3 & B4 & B5 & B6 & B7 & B8 & B9 & B10 & B11 & B12 & B13 \\
	\hline
A1 & 0.008 &  0.018 &  0.022 &  0.024 &  0.022 &  0.018 &  0.016 &  0.012 &  0.009 &  0.006 &  0.004 &  0.002 &  0.001 \\
A2 & 0.024 &  0.055 &  0.065 &  0.072 &  0.068 &  0.054 &  0.049 &  0.037 &  0.027 &  0.017 &  0.011 &  0.006 &  0.003 \\
A3 & 0.020 &  0.044 &  0.053 &  0.059 &  0.055 &  0.044 &  0.040 &  0.030 &  0.022 &  0.014 &  0.009 &  0.005 &  0.003 \\
A4 & 0.024 &  0.055 &  0.066 &  0.073 &  0.068 &  0.055 &  0.049 &  0.037 &  0.027 &  0.017 &  0.011 &  0.006 &  0.003 \\
A5 & 0.024 &  0.054 &  0.065 &  0.072 &  0.067 &  0.054 &  0.049 &  0.036 &  0.026 &  0.017 &  0.011 &  0.006 &  0.003 \\
A6 & 0.020 &  0.046 &  0.055 &  0.061 &  0.057 &  0.046 &  0.041 &  0.031 &  0.023 &  0.014 &  0.009 &  0.005 &  0.003 \\
A7 & 0.020 &  0.045 &  0.054 &  0.060 &  0.056 &  0.045 &  0.041 &  0.030 &  0.022 &  0.014 &  0.009 &  0.005 &  0.003 \\
A8 & 0.019 &  0.042 &  0.051 &  0.056 &  0.052 &  0.042 &  0.038 &  0.028 &  0.021 &  0.013 &  0.009 &  0.005 &  0.003 \\
A9 & 0.020 &  0.045 &  0.055 &  0.060 &  0.056 &  0.045 &  0.041 &  0.031 &  0.022 &  0.014 &  0.009 &  0.005 &  0.003 \\
A10 & 0.023 &  0.051 &  0.062 &  0.068 &  0.064 &  0.051 &  0.046 &  0.035 &  0.025 &  0.016 &  0.010 &  0.006 &  0.003 \\
A11 & 0.022 &  0.049 &  0.058 &  0.064 &  0.060 &  0.048 &  0.043 &  0.033 &  0.024 &  0.015 &  0.010 &  0.006 &  0.003 \\
A12 & 0.019 &  0.043 &  0.052 &  0.057 &  0.053 &  0.043 &  0.039 &  0.029 &  0.021 &  0.013 &  0.009 &  0.005 &  0.003 \\
A13 & -     &  -     &  - &  - &  - &  - &  - &  - &  - & - & - & - &  - \\	
 \hline	
	  \end{tabular}
	  \end{ruledtabular}
    \label{tab:diA_diB}
\end{table*}


\begin{table*}
    \centering
    \caption{The correlation matrix for the point-to-point uncertainties (systematic only) coupling forward-forward dijet measurements (topology A) with central-central dijet measurements (topology C). The relative luminosity and beam polarization uncertainties are not included because they are the same for all points.}
  	\begin{ruledtabular} 
  	\begin{tabular}{c | c c c c c c c c c c c c c c c}
    \hline
    Label & C1 & C2 & C3 & C4 & C5 & C6 & C7 & C8 & C9 & C10 & C11 & C12 & C13 \\
	\hline
A1 & 0.009 &  0.018 &  0.022 &  0.023 &  0.022 &  0.017 &  0.015 &  0.012 &  0.008 &  0.005 &  0.003 &  0.002 &  - \\
A2 & 0.047 &  0.096 &  0.116 &  0.125 &  0.115 &  0.092 &  0.082 &  0.061 &  0.044 &  0.028 &  0.018 &  0.011 &  - \\
A3 & 0.013 &  0.027 &  0.032 &  0.035 &  0.032 &  0.025 &  0.023 &  0.017 &  0.012 &  0.008 &  0.005 &  0.003 &  - \\
A4 & 0.022 &  0.046 &  0.055 &  0.059 &  0.055 &  0.044 &  0.039 &  0.029 &  0.021 &  0.013 &  0.009 &  0.005 &  - \\
A5 & 0.020 &  0.041 &  0.049 &  0.053 &  0.049 &  0.039 &  0.035 &  0.026 &  0.019 &  0.012 &  0.008 &  0.004 &  - \\
A6 & 0.016 &  0.033 &  0.039 &  0.042 &  0.039 &  0.031 &  0.028 &  0.021 &  0.015 &  0.010 &  0.006 &  0.004 &  - \\
A7 & 0.015 &  0.031 &  0.038 &  0.041 &  0.038 &  0.030 &  0.027 &  0.020 &  0.014 &  0.009 &  0.006 &  0.003 &  - \\
A8 & 0.014 &  0.029 &  0.035 &  0.038 &  0.035 &  0.028 &  0.025 &  0.019 &  0.013 &  0.009 &  0.006 &  0.003 &  - \\
A9 & 0.016 &  0.034 &  0.040 &  0.044 &  0.040 &  0.032 &  0.029 &  0.021 &  0.015 &  0.010 &  0.006 &  0.004 &  - \\
A10 & 0.021 &  0.043 &  0.052 &  0.056 &  0.051 &  0.041 &  0.037 &  0.027 &  0.020 &  0.013 &  0.008 &  0.005 & - \\
A11 & 0.020 &  0.040 &  0.049 &  0.052 &  0.048 &  0.039 &  0.035 &  0.026 &  0.019 &  0.012 &  0.008 &  0.004 & - \\
A12 & 0.021 &  0.044 &  0.053 &  0.057 &  0.052 &  0.042 &  0.037 &  0.028 &  0.020 &  0.013 &  0.008 &  0.005 & - \\
A13 & -     &  -     &  - &  - &  - &  - &  - &  - &  - & - & - & - &  - \\	
 \hline	
	  \end{tabular}
	  \end{ruledtabular}
    \label{tab:diA_diC}
\end{table*}


\begin{table*}
    \centering
    \caption{The correlation matrix for the point-to-point uncertainties (systematic only) coupling forward-forward dijet measurements (topology A) with forward-backward dijet measurements (topology D). The relative luminosity and beam polarization uncertainties are not included because they are the same for all points.}
  	\begin{ruledtabular}
  	\begin{tabular}{c | c c c c c c c c c c c c c c c}
    \hline
    Label & D1 & D2 & D3 & D4 & D5 & D6 & D7 & D8 & D9 & D10 & D11 & D12 & D13 \\
	\hline
A1 & - &  0.029 &  0.039 &  0.044 &  0.043 &  0.036 &  0.033 &  0.025 &  0.019 &  0.012 &  0.008 &  0.005 &  0.003 \\
A2 & - &  0.062 &  0.084 &  0.096 &  0.094 &  0.078 &  0.071 &  0.054 &  0.040 &  0.026 &  0.017 &  0.010 &  0.005 \\
A3 & - &  0.026 &  0.035 &  0.041 &  0.040 &  0.033 &  0.030 &  0.023 &  0.017 &  0.011 &  0.007 &  0.004 &  0.002 \\
A4 & - &  0.067 &  0.092 &  0.105 &  0.103 &  0.084 &  0.077 &  0.059 &  0.044 &  0.029 &  0.019 &  0.011 &  0.006 \\
A5 & - &  0.058 &  0.078 &  0.090 &  0.088 &  0.072 &  0.066 &  0.051 &  0.038 &  0.025 &  0.016 &  0.010 &  0.005 \\
A6 & - &  0.043 &  0.058 &  0.067 &  0.065 &  0.054 &  0.049 &  0.038 &  0.028 &  0.018 &  0.012 &  0.007 &  0.004 \\
A7 & - &  0.038 &  0.052 &  0.059 &  0.058 &  0.048 &  0.044 &  0.034 &  0.025 &  0.016 &  0.011 &  0.006 &  0.003 \\
A8 & - &  0.032 &  0.044 &  0.050 &  0.049 &  0.041 &  0.037 &  0.028 &  0.021 &  0.014 &  0.009 &  0.005 &  0.003 \\
A9 & - &  0.031 &  0.042 &  0.048 &  0.047 &  0.038 &  0.035 &  0.027 &  0.020 &  0.013 &  0.009 &  0.005 &  0.003 \\
A10 & - &  0.032 &  0.043 &  0.049 &  0.048 &  0.040 &  0.036 &  0.028 &  0.021 &  0.014 &  0.009 &  0.005 &  0.003 \\
A11 & - &  0.030 &  0.041 &  0.047 &  0.046 &  0.038 &  0.034 &  0.026 &  0.020 &  0.013 &  0.008 &  0.005 &  0.003 \\
A12 & - &  0.028 &  0.038 &  0.044 &  0.043 &  0.035 &  0.032 &  0.025 &  0.018 &  0.012 &  0.008 &  0.005 &  0.002 \\
A13&   -   &    -   &   -    &   -    &    -   &   -    &   -    &    -   &   -    &   -    &    -   &   -    &   -    \\  
 \hline	
	  \end{tabular}
	  \end{ruledtabular}
    \label{tab:diA_diD}
\end{table*}


\begin{table*}
    \centering
    \caption{The correlation matrix for the point-to-point uncertainties (systematic only) for the forward-central dijet measurements (topology B). The relative luminosity and beam polarization uncertainties are not included because they are the same for all points.}
  	\begin{ruledtabular}
  	\begin{tabular}{c | c c c c c c c c c c c c c c c}
    \hline
    Label & B1 & B2 & B3 & B4 & B5 & B6 & B7 & B8 & B9 & B10 & B11 & B12 & B13 \\
	\hline
B1 &   1   &  0.020 &  0.024 &  0.027 &  0.025 &  0.021 &  0.019 &  0.014 &  0.010 &  0.007 &  0.004 &  0.003 &  0.001 \\
B2 &       &    1   &  0.059 &  0.066 &  0.062 &  0.051 &  0.046 &  0.035 &  0.026 &  0.017 &  0.011 &  0.007 &  0.004 \\
B3 &       &        &    1   &  0.080 &  0.076 &  0.061 &  0.056 &  0.042 &  0.031 &  0.020 &  0.013 &  0.008 &  0.004 \\
B4 &       &        &        &    1   &  0.089 &  0.072 &  0.065 &  0.050 &  0.036 &  0.024 &  0.016 &  0.009 &  0.005 \\
B5 &       &        &        &        &    1   &  0.074 &  0.067 &  0.051 &  0.037 &  0.024 &  0.016 &  0.010 &  0.005 \\
B6 &       &        &        &        &        &    1   &  0.058 &  0.044 &  0.033 &  0.021 &  0.014 &  0.008 &  0.004 \\
B7 &       &        &        &        &        &        &    1   &  0.043 &  0.032 &  0.021 &  0.014 &  0.008 &  0.004 \\
B8 &       &        &        &        &        &        &        &    1   &  0.030 &  0.019 &  0.013 &  0.008 &  0.004 \\
B9 &       &        &        &        &        &        &        &        &    1   &  0.021 &  0.014 &  0.008 &  0.004 \\
B10 &       &        &        &        &        &        &        &        &        &    1   &  0.014 &  0.009 &  0.005 \\
B11 &       &        &        &        &        &        &        &        &        &        &    1   &  0.008 &  0.004 \\
B12 &       &        &        &        &        &        &        &        &        &        &        &    1   &  0.003 \\
B13 &       &        &        &        &        &        &        &        &        &        &        &        &    1   \\
 \hline	
	  \end{tabular}
	  \end{ruledtabular}
    \label{tab:diB_diB}
\end{table*}


\begin{table*}
    \centering
    \caption{The correlation matrix for the point-to-point uncertainties (systematic only) coupling forward-central dijet measurements (topology B) with central-central dijet measurements (topology C). The relative luminosity and beam polarization uncertainties are not included because they are the same for all points.}
  	\begin{ruledtabular} 
  	\begin{tabular}{c | c c c c c c c c c c c c c c c}
    \hline
    Label & C1 & C2 & C3 & C4 & C5 & C6 & C7 & C8 & C9 & C10 & C11 & C12 & C13 \\
	\hline
B1 & 0.009 &  0.020 &  0.024 &  0.027 &  0.025 &  0.020 &  0.018 &  0.014 &  0.010 &  0.006 &  0.004 &  0.003 &  - \\
B2 & 0.039 &  0.086 &  0.105 &  0.115 &  0.107 &  0.086 &  0.078 &  0.059 &  0.043 &  0.028 &  0.018 &  0.011 &  - \\
B3 & 0.016 &  0.036 &  0.044 &  0.048 &  0.045 &  0.036 &  0.032 &  0.024 &  0.018 &  0.011 &  0.008 &  0.004 &  - \\
B4 & 0.026 &  0.057 &  0.070 &  0.077 &  0.072 &  0.058 &  0.052 &  0.039 &  0.029 &  0.018 &  0.012 &  0.007 &  - \\
B5 & 0.024 &  0.053 &  0.065 &  0.071 &  0.066 &  0.053 &  0.048 &  0.036 &  0.026 &  0.017 &  0.011 &  0.007 &  - \\
B6 & 0.020 &  0.044 &  0.053 &  0.058 &  0.055 &  0.044 &  0.039 &  0.030 &  0.022 &  0.014 &  0.009 &  0.005 &  - \\
B7 & 0.019 &  0.043 &  0.053 &  0.058 &  0.054 &  0.044 &  0.039 &  0.029 &  0.022 &  0.014 &  0.009 &  0.005 &  - \\
B8 & 0.018 &  0.041 &  0.049 &  0.054 &  0.051 &  0.041 &  0.037 &  0.028 &  0.020 &  0.013 &  0.009 &  0.005 &  - \\
B9 & 0.020 &  0.045 &  0.055 &  0.060 &  0.057 &  0.045 &  0.041 &  0.031 &  0.022 &  0.015 &  0.010 &  0.006 &  - \\
B10 & 0.024 &  0.053 &  0.065 &  0.071 &  0.067 &  0.054 &  0.048 &  0.036 &  0.027 &  0.017 &  0.011 &  0.007 &  - \\
B11 & 0.021 &  0.046 &  0.057 &  0.062 &  0.058 &  0.047 &  0.042 &  0.032 &  0.023 &  0.015 &  0.010 &  0.006 &  - \\
B12 & 0.019 &  0.042 &  0.051 &  0.056 &  0.053 &  0.042 &  0.038 &  0.029 &  0.021 &  0.014 &  0.009 &  0.005 &  - \\
B13 & 0.057 &  0.126 &  0.154 &  0.169 &  0.158 &  0.127 &  0.114 &  0.086 &  0.063 &  0.041 &  0.027 &  0.016 &  - \\	
 \hline	
	  \end{tabular}
	  \end{ruledtabular}
    \label{tab:diB_diC}
\end{table*}


\begin{table*}
    \centering
    \caption{The correlation matrix for the point-to-point uncertainties (systematic only) coupling forward-central dijet measurements (topology B) with forward-backward dijet measurements (topology D). The relative luminosity and beam polarization uncertainties are not included because they are the same for all points.}
  	\begin{ruledtabular} 
  	\begin{tabular}{c | c c c c c c c c c c c c c c c}
    \hline
    Label & D1 & D2 & D3 & D4 & D5 & D6 & D7 & D8 & D9 & D10 & D11 & D12 & D13 \\
	\hline
B1 & - &  0.031 &  0.043 &  0.050 &  0.050 &  0.041 &  0.038 &  0.030 &  0.022 &  0.015 &  0.010 &  0.006 &  0.003 \\
B2 & - &  0.055 &  0.076 &  0.088 &  0.087 &  0.072 &  0.066 &  0.052 &  0.039 &  0.026 &  0.017 &  0.010 &  0.006 \\
B3 & - &  0.035 &  0.048 &  0.055 &  0.055 &  0.046 &  0.042 &  0.033 &  0.025 &  0.016 &  0.011 &  0.007 &  0.004 \\
B4 & - &  0.084 &  0.116 &  0.134 &  0.133 &  0.111 &  0.102 &  0.079 &  0.060 &  0.039 &  0.027 &  0.016 &  0.009 \\
B5 & - &  0.075 &  0.103 &  0.119 &  0.118 &  0.098 &  0.090 &  0.070 &  0.053 &  0.035 &  0.024 &  0.014 &  0.008 \\
B6 & - &  0.057 &  0.079 &  0.091 &  0.090 &  0.075 &  0.069 &  0.054 &  0.040 &  0.027 &  0.018 &  0.011 &  0.006 \\
B7 & - &  0.052 &  0.072 &  0.083 &  0.083 &  0.069 &  0.063 &  0.049 &  0.037 &  0.024 &  0.016 &  0.010 &  0.005 \\
B8 & - &  0.044 &  0.062 &  0.071 &  0.070 &  0.058 &  0.054 &  0.042 &  0.032 &  0.021 &  0.014 &  0.008 &  0.005 \\
B9 & - &  0.041 &  0.057 &  0.065 &  0.065 &  0.054 &  0.050 &  0.039 &  0.029 &  0.019 &  0.013 &  0.008 &  0.004 \\
B10 & - &  0.039 &  0.055 &  0.063 &  0.062 &  0.052 &  0.048 &  0.037 &  0.028 &  0.018 &  0.012 &  0.007 &  0.004 \\
B11 & - &  0.034 &  0.048 &  0.055 &  0.055 &  0.045 &  0.042 &  0.032 &  0.024 &  0.016 &  0.011 &  0.007 &  0.004 \\
B12 & - &  0.027 &  0.037 &  0.043 &  0.043 &  0.035 &  0.033 &  0.025 &  0.019 &  0.013 &  0.008 &  0.005 &  0.003 \\
B13 & - &  0.063 &  0.087 &  0.100 &  0.099 &  0.082 &  0.076 &  0.059 &  0.044 &  0.029 &  0.020 &  0.012 &  0.007 \\	
 \hline	
	  \end{tabular}
	  \end{ruledtabular}
    \label{tab:diB_diD}
\end{table*}


\begin{table*}
    \centering
    \caption{The correlation matrix for the point-to-point uncertainties (systematic only) for the central-central dijet measurements (topology C). The relative luminosity and beam polarization uncertainties are not included because they are the same for all points.}
  	\begin{ruledtabular} 
  	\begin{tabular}{c | c c c c c c c c c c c c c c c}
    \hline
    Label & C1 & C2 & C3 & C4 & C5 & C6 & C7 & C8 & C9 & C10 & C11 & C12 & C13 \\
	\hline
C1 &   1   &  0.020 &  0.025 &  0.026 &  0.024 &  0.019 &  0.017 &  0.013 &  0.009 &  0.006 &  0.004 &  0.002 &  - \\
C2 &       &    1   &  0.183 &  0.196 &  0.181 &  0.145 &  0.130 &  0.097 &  0.070 &  0.045 &  0.030 &  0.018 &  - \\
C3 &       &        &    1   &  0.028 &  0.026 &  0.021 &  0.019 &  0.014 &  0.010 &  0.006 &  0.004 &  0.003 &  - \\
C4 &       &        &        &    1   &  0.058 &  0.046 &  0.041 &  0.031 &  0.022 &  0.014 &  0.009 &  0.006 &  - \\
C5 &       &        &        &        &    1   &  0.038 &  0.034 &  0.025 &  0.018 &  0.012 &  0.008 &  0.005 &  - \\
C6 &       &        &        &        &        &    1   &  0.027 &  0.020 &  0.014 &  0.009 &  0.006 &  0.004 &  - \\
C7 &       &        &        &        &        &        &    1   &  0.019 &  0.014 &  0.009 &  0.006 &  0.004 &  - \\
C8 &       &        &        &        &        &        &        &    1   &  0.013 &  0.008 &  0.006 &  0.003 &  - \\
C9 &       &        &        &        &        &        &        &        &    1   &  0.010 &  0.007 &  0.004 &  - \\
C10 &       &        &        &        &        &        &        &        &        &    1   &  0.009 &  0.005 &  - \\
C11 &       &        &        &        &        &        &        &        &        &        &    1   &  0.005 &  - \\
C12 &       &        &        &        &        &        &        &        &        &        &        &    1   &  - \\
C13 &       &        &        &        &        &        &        &        &        &        &        &        &    -  \\
 \hline	
	  \end{tabular}
	  \end{ruledtabular}
    \label{tab:diC_diC}
\end{table*}


\begin{table*}
    \centering
    \caption{The correlation matrix for the point-to-point uncertainties (systematic only) coupling central-central dijet measurements (topology C) with forward-backward dijet measurements (topology D). The relative luminosity and beam polarization uncertainties are not included because they are the same for all points.}
     \begin{ruledtabular}
     \begin{tabular}{c | c c c c c c c c c c c c c c c}
    \hline
    Label & D1 & D2 & D3 & D4 & D5 & D6 & D7 & D8 & D9 & D10 & D11 & D12 & D13 \\
	\hline

C1 & - &  0.032 &  0.044 &  0.050 &  0.049 &  0.040 &  0.037 &  0.028 &  0.021 &  0.014 &  0.009 &  0.006 &  0.003 \\
C2 & - &  0.097 &  0.134 &  0.152 &  0.150 &  0.123 &  0.113 &  0.087 &  0.065 &  0.043 &  0.029 &  0.017 &  0.010 \\
C3 & - &  0.021 &  0.029 &  0.033 &  0.032 &  0.027 &  0.024 &  0.019 &  0.014 &  0.009 &  0.006 &  0.004 &  0.002 \\
C4 & - &  0.070 &  0.097 &  0.110 &  0.108 &  0.089 &  0.082 &  0.063 &  0.047 &  0.031 &  0.021 &  0.012 &  0.007 \\
C5 & - &  0.056 &  0.078 &  0.088 &  0.087 &  0.071 &  0.065 &  0.050 &  0.037 &  0.025 &  0.017 &  0.010 &  0.006 \\
C6 & - &  0.040 &  0.056 &  0.063 &  0.062 &  0.051 &  0.047 &  0.036 &  0.027 &  0.018 &  0.012 &  0.007 &  0.004 \\
C7 & - &  0.036 &  0.050 &  0.057 &  0.056 &  0.046 &  0.042 &  0.032 &  0.024 &  0.016 &  0.011 &  0.006 &  0.004 \\
C8 & - &  0.031 &  0.043 &  0.049 &  0.048 &  0.039 &  0.036 &  0.028 &  0.021 &  0.014 &  0.009 &  0.006 &  0.003 \\
C9 & - &  0.030 &  0.042 &  0.048 &  0.047 &  0.039 &  0.035 &  0.027 &  0.020 &  0.013 &  0.009 &  0.005 &  0.003 \\
C10 & - &  0.033 &  0.046 &  0.052 &  0.051 &  0.042 &  0.038 &  0.029 &  0.022 &  0.014 &  0.010 &  0.006 &  0.003 \\
C11 & - &  0.029 &  0.040 &  0.045 &  0.044 &  0.036 &  0.033 &  0.026 &  0.019 &  0.013 &  0.008 &  0.005 &  0.003 \\
C12 & - &  0.027 &  0.038 &  0.043 &  0.042 &  0.035 &  0.032 &  0.024 &  0.018 &  0.012 &  0.008 &  0.005 &  0.003 \\
C13&   -   &    -   &   -    &   -    &    -   &   -    &   -    &    -   &   -    &   -    &    -   &   -    &   -    \\ 
 \hline	
	  \end{tabular}
	  \end{ruledtabular}
    \label{tab:diC_diD}
\end{table*}


\begin{table*}
    \centering
    \caption{The correlation matrix for the point-to-point uncertainties (systematic only) for the forward-backward dijet measurements (topology D). The relative luminosity and beam polarization uncertainties are not included because they are the same for all points.}
  	\begin{ruledtabular} 
  	\begin{tabular}{c | c c c c c c c c c c c c c c c}
    \hline
    Label & D1 & D2 & D3 & D4 & D5 & D6 & D7 & D8 & D9 & D10 & D11 & D12 & D13 \\
	\hline
 D1 &   -   &    -   &   -    &   -    &    -   &   -    &   -    &    -   &   -    &   -    &    -   &   -    &   -    \\
 D2 &       &    1   &  0.097 &  0.115 &  0.120 &  0.103 &  0.096 &  0.077 &  0.059 &  0.040 &  0.027 &  0.017 &  0.009 \\
D3 &       &        &    1   &  0.038 &  0.040 &  0.034 &  0.032 &  0.025 &  0.019 &  0.013 &  0.009 &  0.005 &  0.003 \\
D4 &       &        &        &    1   &  0.197 &  0.168 &  0.158 &  0.125 &  0.097 &  0.065 &  0.045 &  0.027 &  0.015 \\
D5 &       &        &        &        &    1   &  0.130 &  0.122 &  0.097 &  0.075 &  0.050 &  0.034 &  0.021 &  0.012 \\
D6 &       &        &        &        &        &    1   &  0.082 &  0.065 &  0.050 &  0.034 &  0.023 &  0.014 &  0.008 \\
D7 &       &        &        &        &        &        &    1   &  0.054 &  0.042 &  0.028 &  0.019 &  0.012 &  0.007 \\
D8 &       &        &        &        &        &        &        &    1   &  0.032 &  0.022 &  0.015 &  0.009 &  0.005 \\
D9 &       &        &        &        &        &        &        &        &    1   &  0.018 &  0.012 &  0.007 &  0.004 \\
D10 &       &        &        &        &        &        &        &        &        &    1   &  0.011 &  0.007 &  0.004 \\
D11 &       &        &        &        &        &        &        &        &        &        &    1   &  0.006 &  0.003 \\
D12 &       &        &        &        &        &        &        &        &        &        &        &    1   &  0.003 \\
D13 &       &        &        &        &        &        &        &        &        &        &        &        &    1   \\  
 \hline	
	  \end{tabular}
	  \end{ruledtabular}
    \label{tab:diD_diD}
\end{table*}

\bibliographystyle{apsrev4-2}
\bibliography{jetALL2013Bib}

\end{document}